%% file: VSCL_NIR-cluster-phot.tex
\RequirePackage{rotating}
\documentclass[fleqn,usenatbib]{mnras}

\usepackage{newtxtext,newtxmath}

\usepackage[T1]{fontenc}

\DeclareRobustCommand{\VAN}[3]{#2}
\let\VANthebibliography\thebibliography
\def\thebibliography{\DeclareRobustCommand{\VAN}[3]{##3}\VANthebibliography}

\usepackage{graphicx}	
\usepackage{amsmath}	
\usepackage[para,online,flushleft]{threeparttable} 
\usepackage{booktabs} 
\usepackage{xtab,booktabs}
\usepackage{subcaption} 
\usepackage{rotating}
\usepackage{lscape}
\usepackage{xcolor}
\usepackage{longtable}
\usepackage{comment}

\input{abb} 

\title[Deep NIR catalogues of the VSCL galaxy clusters]{Galaxy clusters in the Vela supercluster. $-$ I. Deep NIR catalogues}

\author[N. Hatamkhani et al.]{
N. Hatamkhani,$^{1,2}$\thanks{E-mail: narges@saao.ac.za}
R. C. Kraan-Korteweg,$^{1}$, S. L. Blyth$^{1}$, K. Said$^3$ and A. Elagali$^{4,5}$
\\
$^{1}$Department of Astronomy, University of Cape Town, Private Bag X3, Rondebosch 7701, South Africa\\
$^{2}$South African Astronomical Observatory, PO Box 9, 7925 Observatory, Cape Town, South Africa\\
$^3$School of Mathematics and Physics, University of Queensland, Brisbane, QLD 4072, Australia\\
$^4$Minderoo Foundation, 171 - 173 Mounts Bay Road, Perth, WA 6000, Australia\\
$^5$School of Biological Sciences, University of Western Australia, 35 Stirling Highway, Crawley, WA 6009, Australia
}

\date{Accepted XXX. Received YYY; in original form ZZZ}

\pubyear{2023}

\begin{document}
\label{firstpage}
\pagerange{\pageref{firstpage}--\pageref{lastpage}}
\maketitle

\begin{abstract}
We present six deep Near-InfraRed ($JHK_s$) photometric catalogues of  galaxies identified in six cluster candidates (VC02, VC04, VC05, VC08, VC10, VC11) within the Vela Supercluster (VSCL) as part of our efforts to learn more about this large supercluster which extends across the zone of avoidance ($\ell=272\fdg5 \pm 20\deg, b= \pm 10\deg$, at $cz\sim 18000$~km~s$^{-1}$). The observations were conducted with the InfraRed Survey Facility (IRSF), a 1.4m telescope situated at the South African Astronomical Observatory (SAAO) in Sutherland. The images in each cluster cover $\sim80\%$ of their respective  Abell radii. We identified a total number of 1715 galaxies
distributed over the six cluster candidates, of which only $\sim15\%$ were previously known. We study the structures and richnesses of the six clusters out to the cluster-centric completeness radius of $r_c<1.5$~Mpc and magnitude completeness limit of $K_s^o<15\fm5$, using their iso-density contour maps and radial density profiles. The analysis shows VC04 to be the richest of the six. It is  a massive cluster comparable to the Coma and Norma clusters, although its velocity dispersion, $\sigma_v=455~$\kms, seems rather low for a rich cluster. VC02 and VC05 are found to be relatively rich clusters while VC08 is rather poor. Also, VC05 has the highest central number density among the six. VC11 is an intermediate cluster that contains two major subclusters while VC10 has a filament-like structure and is likely not to be a cluster after all.

\end{abstract}




\section{Introduction}
Superclusters are defined as collections of galaxy clusters and groups above a specified spatial overdensity \citep{bahcall1988,einasto2011a,einasto2011b} and contribute to the largest density enhancements in the cosmic web. In 2017 \citet{kraan2017} (hereafter KK2017) uncovered an unknown, relatively nearby ($cz \sim 18000\,$\kms), extended supercluster centred in the region of the Vela constellation ($\ell=272\fdg5 \pm 20\deg, b= \pm 10\deg$). It was identified as part of a long-term program to unveil the large-scale structure of galaxies hidden behind our Milky Way -- the so-called Zone of Avoidance (ZoA).  Spectroscopic follow-ups based on their deep optical galaxy ZoA catalogue \citep[e.g. ][]{kraanlahav2000,kraan2000catalog}, complemented by sources from the 2MASX (2MASS Extended Source) Near-InfraRed (NIR) catalogue  \citep[complete to $K^o_s<13\fm5$;][]{jarrett2000A} in the Vela ZoA region ($245\deg \lesssim \ell \lesssim 295\deg$, $|b|\lesssim 10\deg$), resulted in over 4000 new redshifts being obtained with AAOmega+2dF on the 4-m Australian Telescope, as well as multi-object spectroscopy with the 11-m Southern African Large Telescope (SALT) in the dense cores of potential galaxy clusters \citep[][]{kraan2015salt,kraan2017}. 

The resulting redshift distribution -- although sparsely sampled -- revealed a highly prominent peak at $17000-19000$~\kms, with lower level shoulders extending from $15000-23000$~\kms\ (see Fig.~2 in KK2017), similar in shape to the well studied rich Shapley Supercluster at $14500$~\kms\ \citep[SSC;][]{proust2006}. 
The broad peak in the VSCL velocity range is equally pronounced on each side of the Galactic Plane (GP), suggestive of continuity across the ZoA. As such it would imply an extent of at least $25\deg \times 20\deg$ on the sky, corresponding to $\sim 115 \times 90\, h^{-1}_{70}\,$Mpc at the VSCL distance. A preliminary assessment of the overdensity restricted to two strips ($6\deg <|b|< 10\deg$) where 2MASX is complete, confirmed the galaxy overdensity, both in NIR galaxy counts  as well as within the redshift space surrounding VSCL (using photometric redshifts). The latter would indicate an acceleration on the Local Group ($\sim\,50$\,\kms, see KK2017) of similar order as the SSC \citep{loeb2008}, and is therefore relevant to our understanding of the dynamics in the Local Universe.
 
While hardly any galaxy data exist within $|b|\la 5\deg$ that can trace the VSCL across the inner ZoA, two assessments completely independent of the KK2017 redshift sample were carried out that substantiate its existence. \citet{sorce2017} performed constrained reconstructions (CLUES) using the $CosmicFlows-2$ data set \citep{tully2013}, while \citet{courtois2019} applied a newly derived kinematic model \citep{graziani2019} to the deeper $CosmicFlows-3$ compendium \citep{tully2016} to derive the gravitational effect of the VSCL. Both results support the findings in KK2017, and even find a hint of a hidden core of the VSCL located deep in the ZoA. Taken together this raises the likelihood of the VSCL contributing to the residual bulk flow which originates from distances beyond $\sim 15000$\kms\ in this direction of the sky \citep[see e.g., ][]{springob2016,scrimgeour2016}. Therefore, despite VSCL's location in the ZoA, it is of interest to learn more about the morphology and richness of this partially obscured supercluster.

Massive superclusters are composed of large numbers of galaxy clusters with the larger, more massive superclusters containing increasingly more and richer galaxy clusters \citep{wray2006,einasto2007a}. We therefore decided to investigate the clusters within VSCL in more detail as a first step toward  characterising this partially obscured supercluster. Twenty potential galaxy clusters with velocity dispersions $\sigma _v > 400$\,\kms\ were identified in the VSCL (see Fig.~3 in KK2017). Two of the newly identified potential clusters have X-ray counterparts in ROSAT, although were not recognized as such in the Clusters In the Zone of Avoidance (CIZA) survey, from the ROSAT all-sky survey \citep{ebeling2005}. This would classify the VSCL as a rich supercluster according to \citet{einasto2007a}. In addition, two further VSCL X-ray clusters are identified in CIZA  \citep[CIZA J0812.5-5714 and CIZA J0820.9-5704;][]{kocevski2006} that clearly form part of the VSCL wall although they lie just beyond the borders of the VSCL survey region. 

However, given the distance of the VSCL of $\sim\,260~h^{-1}_{70}~$Mpc, the galaxy catalogue and the redshift follow-ups  that led to its discovery are too shallow for an in-depth investigation of the supercluster's cluster properties. Both the optical and 2MASX selected galaxies are limited to the intrinsically brightest galaxies of the VSCL cluster population ($M_{\rm B^o} \la -19\fm5$ to $-21\fm5$ depending on the foreground extinction, respectively $M_{\rm Ks^o} \la -23\fm5$ in the NIR), reaching only marginally beyond the knee of the luminosity function. This is insufficient to get a better understanding of the richness, morphology and kinematics of these clusters. We therefore conducted deep NIR observations of a selected set of VSCL clusters.

NIR observations are very efficient at probing structures that are affected by dust, be it dust-enshrouded objects, Galactic objects in the Milky Way, or galaxies located behind the Milky Way because of the considerably lower extinction effects in the NIR  compared to the optical (e.g. $A_{K} = 0.09 A_{B}$). Moreover, the NIR is particularly suitable for the detection of early-type galaxies which are dominated by an old stellar component \citep[e.g., ][]{silva2008,cesetti2009}, and predominantly populate the cores of galaxy clusters \citep[morphology-density relation,][]{dressler1980,vogt2004}. 

With this in mind we pursued NIR imaging observations in the $J$, $H$ and $K_s$-bands of ten of the more promising VSCL galaxy clusters. We used the  IRSF telescope located at the SAAO site in Sutherland with the Simultaneous 3-colour Infrared Imager for Unbiased Surveys (SIRIUS) camera. The instrument has excellent resolution (0\farcs45 per pixel) and was shown to  reach a galaxy magnitude $\sim 2\fm0$ deeper than 2MASX in the $K_s$-band with 10~min of observing time. This set-up has successfully been used for various NIR galaxy surveys of clusters  in the southern ZoA  \citep[e.g.,][]{woudt2005,nagayama2006,cluver2008,riad2010} as well as a survey along the Great Attractor Wall \citep{kraan2011}.  As such, it is an ideal instrument for an in-depth study of the VSCL clusters and, by extension, the VSCL itself.

This paper is the first of a series of papers presenting the analysis of deep NIR imaging of six VSCL clusters covering $\sim 80\%$ of their respective Abell radii. It provides a detailed description of the NIR photometry and a first look at the properties of the unveiled galaxies. Subsequent papers will include a derivation of their luminosity functions, and  a more in-depth dynamical analysis of the richest VSCL cluster including new spectroscopy of the fainter galaxies identified in this work.

The structure of this paper is as follows. Section~\ref{sec:nirobs} describes the observations, imaging method and quality assessment of the obtained images. The process of galaxy identification, extraction of their photometric parameters and assessment of the reliability of the derived parameters are explained in Section~\ref{sec:galphot}. The effect of extinction and the $k$-correction on the photometric parameters of galaxies are discussed in Section~\ref{sec:extinction}. The NIR extended source catalogue with the derived parameters is represented in Section~\ref{sec:nircatalog}, as well as a derivation of the completeness magnitude limit of all of the clusters. Section~\ref{sec:sixgalcluster} is dedicated to the analysis of the VSCL galaxy clusters and a summary of the results is given in Section~\ref{sec:summary}.
Throughout this paper we use a cosmological model with parameters: $H_{0}=70\,\mathrm{km~s\textsuperscript{-1}~ Mpc\textsuperscript{-1}}, \Omega_{M}=0.3$ and $\Omega_{\Lambda}=0.7$.

\section{NIR observations}
\label{sec:nirobs}
As mentioned above, the VSCL imaging data were obtained with the IRSF (SAAO), a 1.4m Alt-Azimuth Cassegrain telescope  equipped with the 3-colour NIR imaging camera SIRIUS. SIRIUS utilizes three $1024 \times 1024$ pixel HgCdTe Astronomical Wide Area Infrared Imager (HAWAII) arrays, with a gain of 5.5 e$^{-}$/ADU (Analog-to-Digital Unit) and a readout-noise of 30 e$^{-}$.
The field-of-view of the camera is 7\farcm7$\times$7\farcm7 with a resolution of 0\farcs45 per pixel. An exposure time of 900 seconds results in limiting magnitudes of the SIRIUS camera for point sources of $J = 19\fm0$, $H = 18\fm5$ and  $K_{s} = 17\fm0$ (S/N = $10\,\sigma$) \citep{1999sf99.proc..397N,nagayama2003,nagayama2012}. 

The excellent resolution of IRSF of $0\farcs45$ per pixel, makes it a suitable instrument for surveys of regions with high stellar density. It improves the processes of galaxy identification and star-subtraction due to the better deblending between stars and galaxies, resulting in  more reliable galaxy photometry.

\subsection{Data acquisition}

The observations of the galaxy clusters in the VSCL were performed between 2015 and 2019 during three observing runs. Eight galaxy clusters and two galaxy groups were observed during three weeks in February, March and April 2015 (observers A. Elagali and K. Said). Due to bad weather and a telescope readout issue, more than 60\% of the observed fields in 2015 were not of science quality. A follow-up observing run was performed in April 2017 for two weeks (observers N. Hatamkhani and K. Said), during which the gaps and unsatisfactory fields of the 6 high priority clusters were re-observed. Because of bad weather during the 2017 observing run, almost 40\% of the 150 obtained fields had poor seeing (FWHM$>\!2\farcs2$) and were not suitable for scientific analysis, 
and a third observing run of two weeks was conducted in December 2018 and January 2019 (observer N. Hatamkhani). During this run all the remaining unsatisfactory fields of the six clusters could be re-observed. 
This resulted in a final tally of six clusters for which we achieved complete coverage over the envisioned cluster survey area ($\sim 2/3 R_A$) that had good science quality.

The final six clusters are listed in Table~\ref{tab:clusterstat} with an internal cluster ID, their central positions (equatorial), mean recession velocity and velocity dispersion. The number of galaxies with known redshifts including foreground and background galaxies ($N_r$), and galaxies at the distance of each cluster candidate over the surveyed area ($N_m$) are given in the last two columns of the table respectively.

\begin{table}
\caption{Properties of the six galaxy clusters within the VSCL selected for a NIR study (based on the KK2017 redshift catalogue). $cz$ and $\sigma_v$ denote the mean recession velocity and velocity dispersion respectively.  $N_r$ is the total number of galaxies with known redshifts, including foreground and background galaxies. $N_m$ is the number of galaxies at the distance of each cluster candidate.}
\begin{center}
\begin{threeparttable}
\scalebox{0.95}{
\begin{tabular}{cccccc}
\toprule
\toprule
 Name& Center ($\alpha$, $\delta$) & $cz$ & $\sigma_v$ & $N_r$ & $N_m$ \\
  & (deg) &km~s$^{-1}$ & km~s$^{-1}$ &  &  \\
 \midrule
 VC02 &122.69, $-49.39$  &18985 &659 & 82 & 56 \\
 VC04 &129.14, $-55.81$ &18196 &455 & 89 & 85 \\
 VC05 &129.78, $-57.47$ &21275 &469 & 52 & 30 \\
 VC08 &134.54, $-57.75$ &17316 &355 & 38 & 36 \\
 VC10 &147.38, $-43.84$ &18173 &428 & 29 & 27 \\
 VC11 &149.00, $-43.89$  &17919 &633 & 67  & 65 \\
 \bottomrule
\end{tabular}}
\end{threeparttable}
\end{center}
\label{tab:clusterstat}
\end{table}

\subsection{Imaging method}
\label{subsubsec:imaging}
\begin{figure}
\centering
\includegraphics[width=\columnwidth, height=0.25\textheight, keepaspectratio]{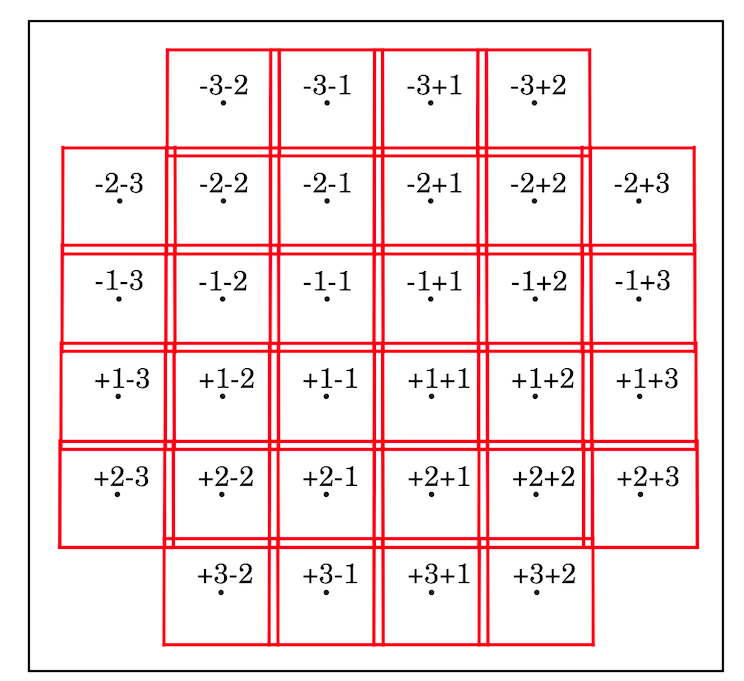}
\caption{A sample of the observed mosaic consists of 32 slightly overlapping fields for each cluster. In the notation $\pm a\pm b$, $a$ and $b$ are the numerical offsets in the horizontal and vertical directions with respect to the cluster center, used to identify a field within a mosaic.}
\label{fig:mosaic}
\end{figure}

For a cluster at the VSCL redshift of $cz \sim$18000~km~s$^{-1}$, the Abell radius ($R_A$) corresponds to 28\farcm6 which translates to $\sim 2.1~h^{-1}_{70}$~Mpc.
Given the IRSF Field-of-View (FoV) of 7\farcm7 versus the Abell radius of 28\farcm6, a mosaic of 32 fields was defined to provide good coverage of the cluster cores ($r\la 0.7 R_A$). Each cluster consists of 4 rows of 6 slightly overlapping fields with an additional row of 4 fields at the top and bottom, thus bringing the total number of observed fields  to 32 (Fig.~\ref{fig:mosaic}). Using this method of observing, each cluster mosaic covers almost 80\% of the Abell radius and therefore, the completeness radius of this survey varies from $1.5-1.8~$Mpc depending on the exact distance of the observed cluster. 

We used a dithering method to remove the faulty pixels in the NIR detector. Each field was observed for  25 successive frames of 24s (600s in total) and as a result this study has limiting magnitude $\sim $2\fm0  deeper than the 2MASX survey \citep{jarrett2000}. Observing for longer than 10 minutes in the densely populated ZoA would result in increased sky background noise due to NIR emission from the old stellar population in the Galaxy and will impact the reliability of photometry \citep{riad2010}.

\subsection{Data reduction and image quality assessment}
\label{subsec:reduction} 
The SIRIUS pipeline, developed and maintained by Yasushi Nakajima, was used to perform a preliminary data reduction on all the raw frames obtained with the SIRIUS camera. The image processing consists of dark frame subtraction, flat-fielding, sky determination and sky subtraction, dither combination, astrometric and photometric calibrations \cite[see][and the SIRIUS pipeline manual\footnote{\url{https://sourceforge.net/projects/irsfsoftware/files/}} for more details]{riad2010,williams2014}.

To ensure that all the images were observed under photometric conditions, their seeing values  and magnitude zero-points (ZPs) in the three observing periods were assessed. In total there are 263 images distributed over 6 mosaics in this survey. The values of the average seeing and the mean ZPs with their respective standard deviations ($\sigma$) are given in Table~\ref{tab:imagestat}. The images used for further analysis are restricted to have ZPs that lie within a value of $\langle ZP \rangle \pm 2\sigma$ and a seeing of FWHM$<2\farcs3$.

\begin{table}
\caption{Mean magnitude zero-point, $\langle ZP \rangle$, and average seeing with their respective standard deviations, $\sigma$, for all 263 images.}
\begin{center}
\scalebox{0.95}{
\begin{tabular}{ccccc}
\toprule
\toprule
 Band & $\langle ZP \rangle$  & $\sigma$  & Seeing  & $\sigma$ \\
   & (mag)& (mag) & (arcsec) & (arcsec)\\
 \midrule
 $J$ & 20.604 & 0.185& 1.55& 0.316 \\
 $H$ & 20.773 & 0.172& 1.46& 0.302 \\
 $K_{s}$ & 19.967 & 0.167& 1.37& 0.265 \\

 \bottomrule
\end{tabular}}
\end{center}
\label{tab:imagestat}
\end{table}

\section{Galaxy Photometry}
\label{sec:galphot}
After the preliminary data reduction using the SIRIUS pipeline, the images were ready for extraction of sources to create a catalogue of galaxies with their photometric parameters. 
\subsection{Galaxy identification}

Various software packages such as \textsc{focas} \citep{jarvis1981}, \textsc{cosmos} \citep{beard1990} and Source Extractor \citep{bertin1996}, have been developed to automate the process of identifying galaxies, stars or other astronomical objects. The source identification routines in these algorithms perform well for data at high Galactic latitudes, however their performance deteriorates for sources in crowded fields with high foreground extinction, as for instance in the ZoA \citep{beard1990}. However, manual source identification is time consuming and can have observer-dependant biases, i.e. the number of identified galaxies can vary from one person to another. We therefore opted for a combination of automatic and manual galaxy identification, by using Source Extractor to identify the potential sources, then visually inspecting and verifying each identified source using the RGB images in a pipeline specifically developed for analysing IRSF data \citep[hereafter the IRSF pipeline, ][]{williams2014}.

\subsubsection{Visual inspection in a sample of fields}
\label{sec:sampleanalysis}
\begin{figure}
\centering
\includegraphics[width=1\columnwidth, height=\textheight, keepaspectratio]{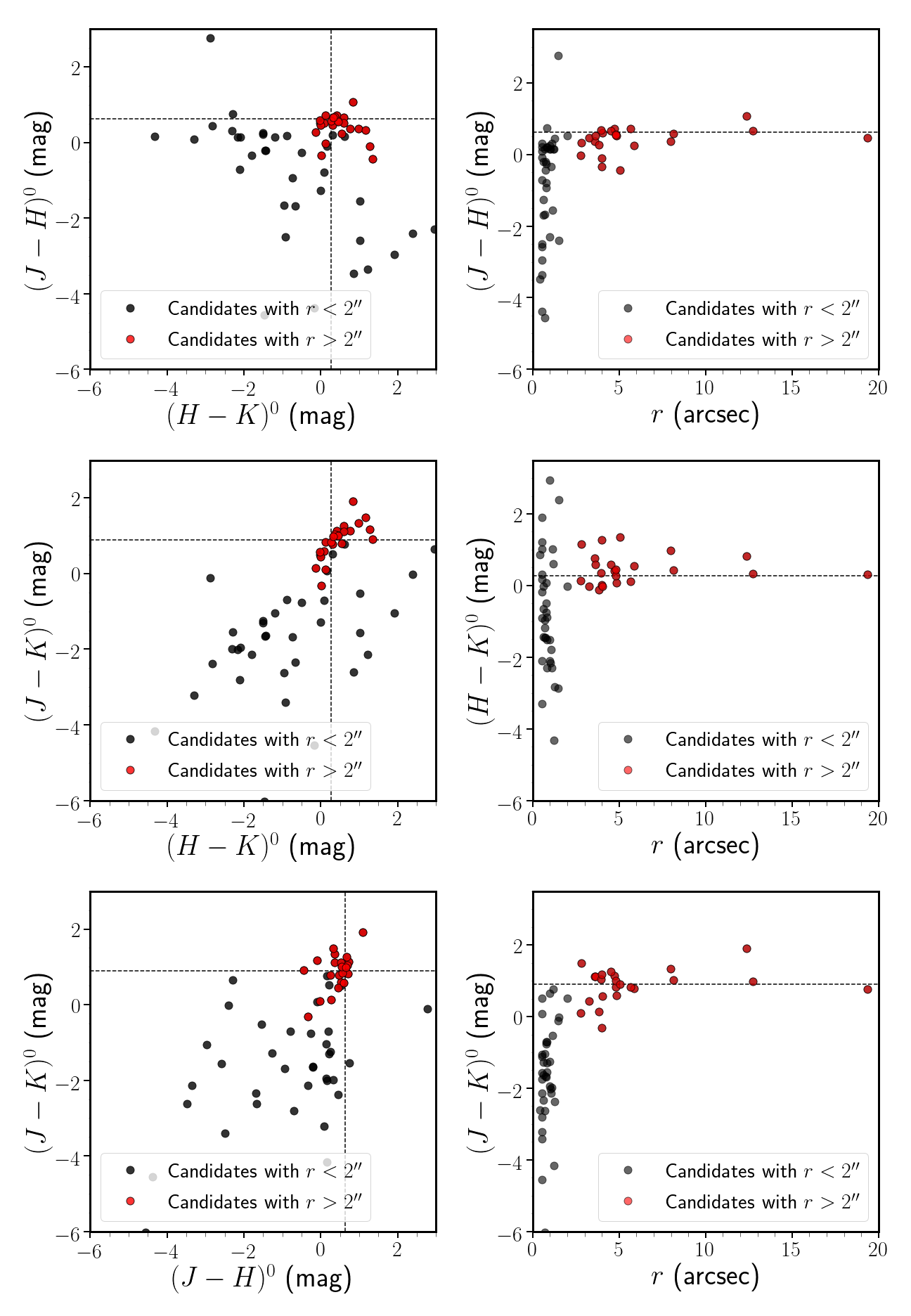}

\caption{Colour-colour (left panel) and colour-radius (right panel) diagrams of 68 galaxy candidates in nine fields of VC08.}
\label{fig:colour-radius}
\end{figure}

Before utilizing any software to identify galaxies, we started by processing a sample of nine IRSF fields in VC08 to get familiarised with the way obscured galaxies in the ZoA can appear on the IRSF images. We first visually identified 68 potential galaxies in RGB composite images. Next, we ran the IRSF pipeline (discussed in detail in Section~\ref{sec:irsfpipeline}) on these candidates to derive their photometric parameters such as the isophotal $J$, $H$ and $K_{s}$ magnitudes. 
To differentiate between true galaxies and other extended Galactic objects or blended stars, we used various diagnostics such as colour-colour and colour-radius diagrams, in addition to visual inspection of these objects. In Fig.~\ref{fig:colour-radius} we show the extinction-corrected colour-colour and colour-radius diagrams in the left and right columns respectively for 68 visually identified galaxies in the nine fields.

The dashed lines in the diagrams show the mean NIR isophotal colours determined by \citet{jarrett2000b-k}, $(J-H)$=0\fm64, $(H-K_{s})$=0\fm28 and $(J-K_{s})$=0\fm92 for nearby (low redshift) early-type galaxies. The candidate galaxies with isophotal radius $r\geqslant 2''$ are shown in red\footnote{The isophotal radius, $r$, is the radius of the isophote at a surface brightness of 20~mag~arcsec$^{-2}$.}. The cut-off radius was selected after analysing the colour-radius diagrams which showed a clear gap around $r=2''$: the objects with $r<2''$ have random NIR colours around the mean colour values of galaxies. Moreover, the colour-colour diagrams revealed that objects with $r<2''$ mostly have bluer colours which indicates that they are either stars or other Galactic objects but not galaxies.

\subsubsection{Source Extractor}
We used Source Extractor program \citep{bertin1996} to identify galaxies in the VSCL survey. Although Source Extractor is particularly oriented towards the reduction of large scale galaxy-survey data (located generally at latitudes far from the Milky Way), it performs reasonably well on moderately crowded star fields \citep{bertin1996}. The Source Extractor image analysis steps are explained in detail in \citet{bertin1996}.  

Source Extractor uses a Neural Network (NN) to classify the extracted objects either as stars or extended sources. The NN yields a parameter, CLASS-STAR, that indicates the stellarity-index. We use only those sources for which Class-Star $<0.35$, i.e. classified as galaxies. 
The Source Extractor configuration file must be adjusted to the characteristics of the imaged fields by the user. Since the VSCL survey was carried out in the ZoA, the high stellar density makes it quite hard for Source Extractor to deblend adjacent stars and other Galactic objects and the configuration parameters needed to be optimised for this work. We therefore, first performed a deep visual search for galaxies on the sample of nine fields from VC08 (Section \ref{sec:sampleanalysis}). The Source Extractor parameters were then chosen empirically and iterated on to arrive at a catalogue that included all the objects that were visually confirmed as galaxies. To minimise the number of false positives, the Source Extractor galaxy output list objects were all visually inspected. A sample configuration file with optimized parameter values for the VSCL survey images is provided in Table~\ref{tab:Source Extractor}.
Source Extractor was then run on all of the IRSF fields in the six VSCL cluster mosaics and the required output catalogues of identified sources were generated as input to run the IRSF photometry pipeline.

\subsection{The IRSF pipeline}
\label{sec:irsfpipeline}

\subsubsection{Galaxy verification}
\begin{figure}
\centering
\includegraphics[width=\columnwidth, height=\textheight, keepaspectratio]{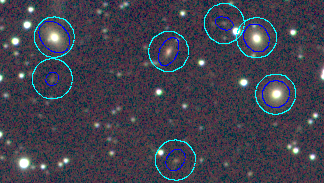}
\caption{An example of how \texttt{findgal} highlights and displays potential candidates in a field. Light blue circles indicate the galaxy candidates. The dark blue ellipses are drawn based on the first estimation of the minor and major axes by the user. The size of this postage stamp is $\sim 2\farcs4 \times 1\farcs3$.}
\label{fig:ds9}
\end{figure}

We used the script \texttt{findgal} developed by \citet{williams2011} to verify that the objects identified by Source Extractor were in fact galaxies.  
The script displays RGB composite images with the interactive \texttt{DS9} interface with which the user can mark the approximate centers of the galaxies and specify their major and minor axes.  This information  is recorded in the form of a table and is subsequently used in the pipeline that performs the star subtraction and photometry. Fig.~\ref{fig:ds9} shows an example of objects found by Source Extractor and displayed by \texttt{DS9} using the script \texttt{findgal}. These sources were then saved for processing through the rest of the IRSF photometry pipeline.

\subsubsection{Star subtraction}
In the regions close to the GP $(|b|\leq 10\deg$), increased stellar density results in higher contamination from neighbouring or overlapping stars \citep{jarrett2000A}. This has two consequences for galaxy photometry: isolating the flux of a galaxy becomes more difficult due to the light from nearby or overlapping sources. Moreover, the flux from unresolved stars in crowded fields causes an increase in the noise amplitude of the sky background (or background surface brightness). This  makes the galaxies appear smaller than their actual sizes since the outer regions of the Low Surface Brightness (LSB) galaxies fade into the background sky, and complicates the detection of LSB galaxies overall.

The second part of the pipeline removes the stars around the detected galaxies via Point-Spread-Function (PSF) fitting. It first determines an average PSF for point sources in the field which is used to remove the stars in a box around each detected galaxy.  The PSF-fitting routine was developed by Takahiro Nagayama (private communication) and \citet{riad2010} using the \texttt{DAOPHOT} package in \texttt{IRAF}, and was modified by \citet{williams2011} to run in \texttt{PyRAF}\footnote{\url{http://www.stsci.edu/institute/software_hardware/pyraf}}. The star-subtraction script performs the following steps: (a) In each image the sky background and the root mean square (rms), $\sigma$, is determined using the \texttt{imstatistics} \texttt{IRAF} task with 30 iterations and $3\sigma$ clipping; (b) the galaxy is modeled using the \texttt{ellipse} and \texttt{bmodel} tasks and subtracted using the \texttt{imarith} task; (c) the bright stars are detected in the galaxy-subtracted image and removed. Source Extractor is used to detect all the sources above $3\sigma$ of the background. PSF photometry and removal of these sources is done by the \texttt{phot} and \texttt{allstar} tasks in the \textsc{daophot} package of \texttt{IRAF}. A file is created by \texttt{allstar} containing the details of the subtracted stars; (d) the previous procedure is repeated to detect and remove faint stars, with the threshold parameter of Source Extractor chosen to be $1.8\sigma$ of the sky background and another list of subtracted stars is created and appended to the original list. All these stars are removed from the original image; (e) any flaw in the PSF-fitting results in residuals in the image, are then removed; (f) steps (b) to (e) are repeated four times. Each iteration improves the PSF model leading to more accurate galaxy photometry.

\begin{figure} 
\centering
\includegraphics[width=1.2\columnwidth, height=0.4\textheight, keepaspectratio]{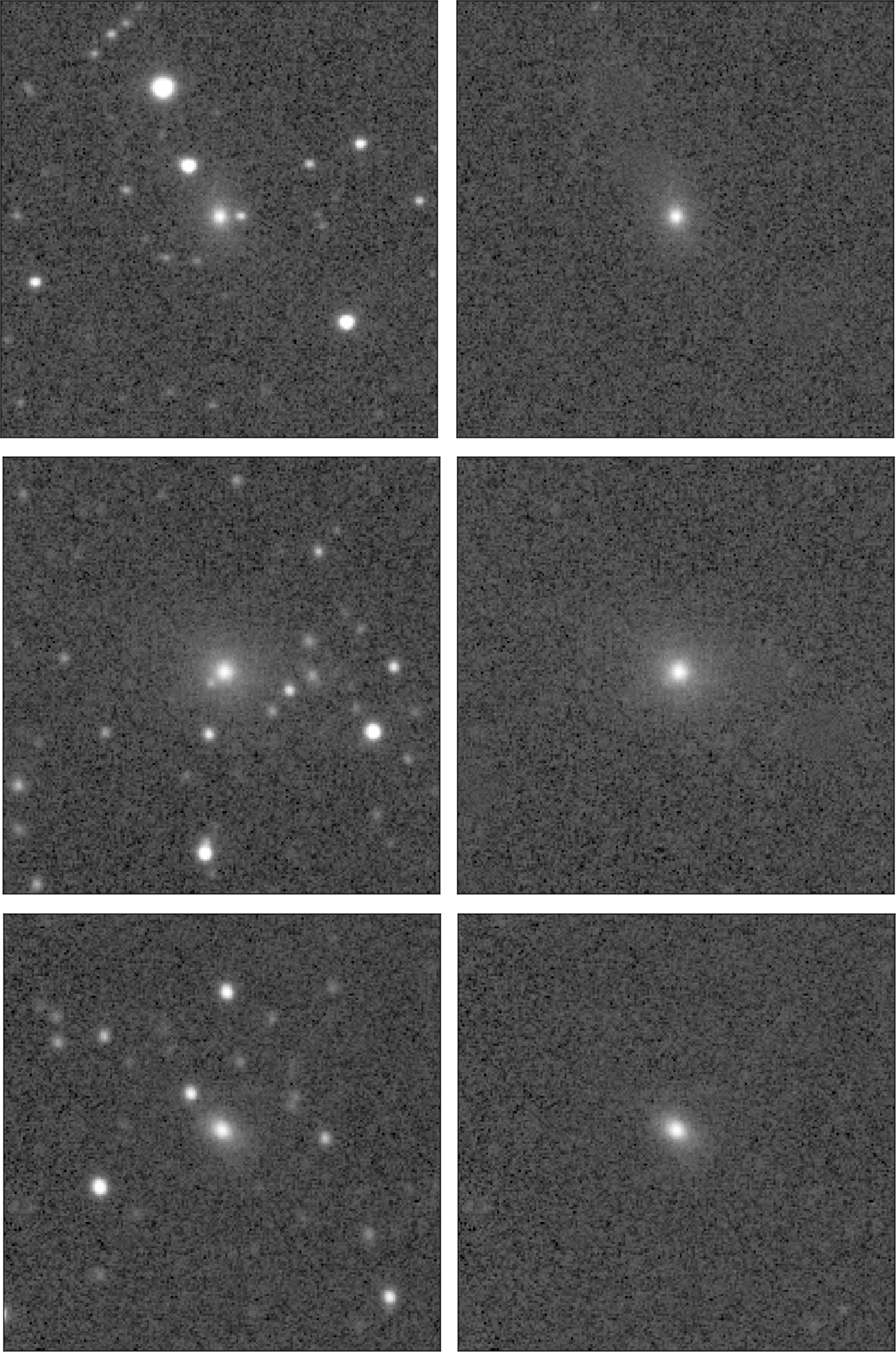}
\caption{The postage stamps of three galaxy candidates in the VC08 cluster at $\langle v \rangle \sim 17316~$km~s$^{-1}$ in the $K_{s}$-band, before (left) and after (right) star-subtraction. The size of the cutouts is $200\times 200$ pixels which translate to $90''\times 90''$ given the resolution of the IRSF of $0\farcs45$ pix\textsuperscript{-1}.}
\label{fig:starsub}
\end{figure}

Fig.~\ref{fig:starsub} shows the cutouts of three galaxy candidates in VC08 before and after star-subtraction.  \citet{nagayama2004} showed, using simulations, that the photometry after  star-subtraction process in star-crowded regions led to only a small systematic difference of $\Delta m=0\fm04 \pm 0\fm2$ compared to their photometry in regions without star-crowding. The difference is well within the standard deviation of the derived magnitudes in the VSCL survey. This means that the star-subtraction process works satisfactorily in these star-crowded regions.

\subsubsection{Galaxy photometry}
The third part of the pipeline extracts astrometric, geometric and photometric parameters of the galaxies. Some of the derived parameters are explained below.

The central positions are determined from the intensity-weighted centroid of the the $J$ + $H$ + $K_{s}$ `super co-add' images.

The \texttt{IRAF} task \texttt{ellipse} is used to determine the ellipticity ($\epsilon=1-b/a$) and position angle ($\phi$, measured counterclockwise from +Y), for a fixed $(X,Y)$ coordinate by fitting isophotes to the galaxy images in each band individually. Each isophote is fitted at a pre-defined, fixed semi-major axis length. The first elliptical isophote is based on an estimate using the approximate values of the center coordinates (X,Y), ellipticity and position angle of the galaxy. This reduces the two-dimensional image into a one-dimensional intensity distribution as a function of position angle.
The ellipticity and position angle are measured in the outer parts of the disk, where they are usually stable. The average value of the ellipticity and position angle is determined between the $1\sigma$ and $2\sigma$ isophotes, where $\sigma$ is the sky rms.

The astrometric and geometric parameters obtained above are used to compute the one-dimensional radial surface brightness profiles (SBP) of the galaxy candidates. 
The pipeline then uses a double S\'{e}rsic function to simultaneously fit an inner bulge and an outer disk to the galaxies.

An estimation of the remaining undetected flux below the sky background noise can be obtained by extrapolating the double S\'{e}rsic function to infinity, following \citet{kirby2008}. The total magnitude is determined as 
\begin{equation}
\begin{aligned}
m_{\rm tot}=m_{\rm obs}-\Delta m\quad,
\end{aligned}
\label{eq:totalmag4}
\end{equation}
where $m_{\rm {obs}}$ is the magnitude contained within the maximum radius (the radius at which surface brightness drops to $1\sigma$ of the sky background) of the galaxy and $\Delta m$ is its magnitude between the maximum radius and the limit $r=\infty$.

The isophotal magnitude, $m_{\mu}$, is measured within a given radius or aperture, using the \texttt{ellipse} task. 
Both $m_{\mu}$ and $r_{\mu}$ are determined at the points $\mu = 20$~mag~arcsec$^{-2}$, $\mu = 21$~mag~arcsec$^{-2}$ and $\mu = 22$~mag~arcsec$^{-2}$ in the surface brightness profile. The magnitudes are derived in the Vega system.

\subsection{Quality control}
We performed several quality checks to ensure the accuracy of the derived astrometric and photometric  parameters as detailed below.

\subsubsection{Internal consistency}
Galaxies close to the border or a corner of an image can be identified in up to three images within a cluster mosaic  (see Section~\ref{subsubsec:imaging}).  We used galaxies in these overlap regions  to check the consistency of astrometric and photometric parameters.

Table~\ref{tab:deltaradec2} compares the mean values of $\Delta$RA and $\Delta$Dec and their standard deviations for the twice-observed galaxies in the six VSCL clusters with those of the Norma Wall Survey which was done with the same telescope, instrument and observation set-up \citep [NWS;][]{riad2010}. The values of $\langle \Delta$RA$\rangle$ and $\langle \Delta$Dec$\rangle$ in each cluster are insignificant and the small offsets are likely due to the differences in the quality of images (e.g. different seeing).
The positional offsets are on average slightly larger than the results from the analysis of the NWS data (e.g. VC04 with $\langle \Delta$RA$\rangle = 0.031$ compared to NWS with $\langle \Delta$RA$\rangle = 0.009$). However, they are consistent when the uncertainties are considered. There is no systematic shift seen in the position of galaxies in each cluster. 

\begin{table}
\caption{$\langle \Delta$RA$\rangle$ and $ \langle \Delta$Dec$\rangle$ with their respective standard deviation ($\sigma$) for the matched pairs in  NWS and the VSCL clusters. $N$ is the number of twice-observed galaxies.}
\begin{center}
\scalebox{0.95}{
\begin{tabular}{crcrcc}
\toprule
\toprule
 Survey & $\langle \Delta$RA$\rangle$ & $\sigma$ & $\langle \Delta$Dec$\rangle$ & $\sigma$& $N$ \\
  & (arcsec) & (arcsec) & (arcsec) & (arcsec) &  \\
 \midrule
 VC02 & $-0.011$ &$0.492$ & $0.023$ & $0.316$& 124 \\
 VC04 & $0.031$ &$0.537$ & $0.046$ & $0.318$ & 249 \\
 VC05 & $0.021$ &$0.754$ & $0.039$ & $0.395$& 121 \\
 VC08 & $-0.011$ &$0.253$ & $0.071$ & $0.117$& 4 \\
 VC10 & $-0.106$ &$0.649$ & $-0.063$ & $0.410$& 43 \\
 VC11 & $0.060$ &$0.507$ & $0.028$ & $0.404$& 78 \\
 \midrule
 NWS  & $0.009$ & $0.202$ & $0.003$ & $0.202$& 1061 \\
 
 \bottomrule
\end{tabular}}
\end{center}
\label{tab:deltaradec2}
\end{table}

To check the consistency of the photometric parameters, we compared the $K_{s20}$ fiducial isophotal magnitudes for each galaxy observed in two different fields by computing:  
 \begin{equation}
\begin{aligned}
\Delta m=m_{\mathrm{field}}-m_{\mathrm{neighbouring\, field}}.
\end{aligned}
\label{eq:shiftinternmag}
\end{equation}

The values of the mean magnitude difference and the standard deviation of VC04 and the NWS \citep{riad2010} are given in Table~\ref{tab:interncomp}. The results of VC04 are typical of what has been found in the other observed clusters. The mean magnitude difference and standard deviation are consistent between the two fields and show no systematic differences. The offsets in VC04 are within $1\sigma$ of those of the NWS ($\langle \Delta J \rangle = 0.012\pm 0.161$, $\langle \Delta H \rangle = 0.003\pm 0.156$ and $\langle \Delta K_s \rangle = 0.006\pm 0.143$).
Note that the standard deviation values in both positions and magnitudes are high because they are derived for the galaxies located at the outskirts of the images.

\begin{table}
\caption{The values of the mean magnitude difference and the standard deviation of one cluster in this survey (VC04 with 249 matched pairs) and the NWS (1061 matched pairs) \citep{riad2010}.}
\begin{center}
\scalebox{0.95}{
\begin{tabular}{crccc}
\toprule
\toprule
 Band  & $\langle \Delta m \rangle$ & $\sigma$ & $\langle \Delta m \rangle$ & $\sigma$ \\
   &VC04 &VC04 &NWS &NWS \\
   & (mag)&(mag)& (mag) & (mag)\\
 \midrule
 $J$ & $0.014$ &$0.169$& $0.012$ &$0.161$ \\
 $H$ &  $0.007$ &$0.181$& $0.003$&$0.156$ \\
 $K_{s}$ & $0.025$ &$0.180$& $0.006$ &$0.143$ \\

 \bottomrule
\end{tabular}}
\end{center}
\label{tab:interncomp}
\end{table}

\subsubsection{External consistency}

\subsubsection*{Astrometric parameters}
 
As a further consistency check we compared our extracted galaxy parameters with their respective 2MASX \citep{jarrett2000} properties. Any two candidates in the catalogues (the VSCL clusters and 2MASX) that lie within a distance of r$\,\leq$\,2\farcs0 from each other were assumed to be counterparts.

The astrometric comparison was done by computing the offsets $\Delta$RA and $\Delta$Dec for each matched pair. It should be noted that the determination of the positions are not derived in exactly the same manner: 2MASX uses the pixel with the peak $J$-band flux as the centre of a galaxy to perform photometry while our method is based on the centroid of the co-added image.  The $\langle \Delta $RA$\rangle$ and $\langle \Delta$Dec$\rangle$ offsets with their dispersions are presented in Table~\ref{tab:deltaradec}. The positions of the VSCL galaxies are consistent with 2MASX and no systematic offsets are seen.  The 2MASX coordinates have a rms uncertainty of $\sim 0\farcs5$ \citep{jarrett2000A} which is compatible with what is found for the VSCL clusters, despite the differences in the determination methods and other survey characteristics such as the pixel size and seeing of the observed images. The offsets were also compared to that of \citet{riad2010} for the Norma wall. The offsets between the NWS 237 cross-matched galaxies are listed in Table~\ref{tab:deltaradec}. They agree with those obtained for the VSCL and 2MASX galaxies. 

\begin{table}
\caption{$\langle \Delta$RA$\rangle$ and $\langle \Delta$Dec$\rangle$ with their respective standard deviations ($\sigma$) for the matched pairs in VC02, VC04, VC05, VC08, VC10, VC11 and NWS respectively. $N_{2MASX}$ is the number of galaxies with 2MASX counterpart in each cluster.}
\begin{center}
\scalebox{0.95}{
\begin{tabular}{crcrcc}
\toprule
\toprule
 Survey & $\langle \Delta$RA$\rangle$ & $\sigma$ & $\langle \Delta$Dec$\rangle$ & $\sigma$& $N_{2MASX}$ \\
  & (arcsec) & (arcsec) & (arcsec) & (arcsec) & \\
 \midrule
 VC02 & $-0.089$ & $0.570$ & $0.045$ &  $0.383$& 43 \\
 VC04 & $-0.055$ & $0.749$ & $0.122$ &  $0.502$& 82 \\
 VC05 & $-0.078$ & $0.628$ & $0.245$ &  $0.451$& 35 \\
 VC08 & $0.086$ & $0.695$ & $0.023$ &  $0.495$& 31 \\
 VC10 & $0.071$ &$0.464$ & $-0.047$ &  $0.468$& 29 \\
 VC11 & $0.048$ & $0.592$ & $-0.083$ &  $0.402$& 48 \\
 \midrule
 NWS & $-0.010$ & $0.420$ & $0.110$&  $0.390$& 237 \\

 \bottomrule
\end{tabular}}
\end{center}
\label{tab:deltaradec}
\end{table}

\subsubsection*{Photometric Parameters}

To verify the accuracy of the IRSF/SIRIUS magnitude system, we computed the difference between the IRSF/SIRIUS and 2MASX $K_{s20}$ fiducial isophotal and $7''$ aperture magnitudes of the 82 cross-identified galaxies in the VC04 cluster, in the $J$, $H$ and $K_{s}$-bands using the metric:
\begin{equation}
\begin{aligned}
\Delta m=m_{IRSF}-m_{2MASS}.
\end{aligned}
\label{eq:shiftmag}
\end{equation}
A positive value of $\Delta m$ would indicate that the magnitudes of the IRSF/SIRIUS system are fainter.

Before performing the photometric comparison and calculating the magnitude differences, we converted 2MASX magnitudes to the SIRIUS filter magnitude system using Equation~\ref{eq:magconv}, provided by Yasushi Nakajima (private communications):

\footnotesize{\begin{equation}
\begin{aligned}
J_{\textsc{sirius}}&= J_{\textsc{2mass}}+(-0.045\pm0.008)( J- H)_{\textsc{2mass}}+(-0.001\pm0.008)\\
H_{\textsc{sirius}}&= H_{\textsc{2mass}}+(0.027\pm0.007)( J- H)_{\textsc{2mass}}+(-0.009\pm0.008)\\
K_{s_{\textsc{sirius}}}&= K_{s_{\textsc{2mass}}}+(0.015\pm0.008)( J- K_{s})_{\textsc{2mass}}+(-0.001\pm0.008).\\
\end{aligned}
\label{eq:magconv}
\end{equation}}\normalsize

Table~\ref{tab:comp7,20}
lists the offset values with their standard deviations for both $7''$ aperture and $K_{s20}$ fiducial isophotal magnitudes. The offsets and dispersions of the $7''$ aperture magnitudes are, as expected, smaller compared to the values for the $K_{s20}$ fiducial isophotal magnitudes.

\begin{table}
\caption{Comparison of the 2MASX and IRSF/SIRIUS $7''$ and $K_{s20}$ fiducial isophotal magnitudes and their respective standard deviations for 82 matched pairs in VC04.}
\begin{center}
\scalebox{0.95}{
\begin{tabular}{ccc}
\toprule
\toprule
Band & $7''$ aperture & $K_{s20}$ fiducial \\
 & $\Delta m$ (mag) & $\Delta m$ (mag)  \\
 \midrule
 $J$ & $0.064 \pm 0.168$ & $0.118 \pm 0.177$  \\
 $H$ & $0.035 \pm 0.123$ & $0.064 \pm 0.167$\\
 $K_{s}$ & $0.053 \pm 0.148$ & $0.091 \pm 0.196$\\

 \bottomrule
\end{tabular}}
\end{center}
\label{tab:comp7,20}
\end{table}

Table~\ref{tab:compk20} lists the offsets and dispersions of the $K_{s20}$ fiducial isophotal magnitudes to compare  them with other surveys pursued with IRSF/SIRIUS that used the same imaging set up \citep{cluver2008,riad2010,williams2014}. The VC04 offsets are within $1\sigma$ of the findings of the other surveys. The increase in $\Delta m$ at the faint end, particularly for the $K_{s20}$ fiducial isophotal magnitudes, could be due to the increase in 2MASX errors for the low surface brightness galaxies (due to losing their flux at their outer areas) and also the lower spatial resolution of 2MASX which could result in not resolving the stars that are superimposed on galaxies (although star-subtraction was a part of the 2MASX photometry of galaxies) hence resulting in brighter magnitudes.   

\begin{table}
\caption{Comparison of the IRSF/SIRIUS $K_{s20}$ fiducial isophotal magnitude offsets with other surveys in $J$, $H$ and $K_{s}$-bands.}
\begin{center}
\scalebox{0.85}{
\begin{tabular}{cccrr}
\toprule
\toprule
 Band & $\Delta m_{K20}$ (mag) & $\Delta m_{K20}$ (mag)  & $\Delta m_{K20}$ (mag)  & $\Delta m_{K20}$ (mag) \\
 & VC04  & \cite{williams2011} & \cite{riad2010}  & \cite{cluver2008}  \\
 \midrule
 $J$ & $0.118 \pm 0.177$ & $-0.070 \pm 0.190$& $-0.029 \pm 0.098$ & $-0.070 \pm 0.110$ \\
 $H$ & $0.064 \pm 0.167$ & $-0.030 \pm 0.170$& $0.051 \pm 0.101$ &$0.030 \pm 0.120$\\
 $K_{s}$ & $0.091 \pm 0.196$& $-0.030 \pm 0.180$& $0.045 \pm 0.095$ &$0.020 \pm 0.070$\\

 \bottomrule
\end{tabular}}
\end{center}
\label{tab:compk20}
\end{table}

\section{Accounting for extinction effects}
\label{sec:extinction}
\subsection{Foreground extinction}
The light of the extended objects in the ZoA is strongly absorbed by the dust of the Milky Way. The Galactic interstellar extinction depends on the wavelength of the light and the relative size of the dust grains. Extinction increases for shorter wavelengths and therefore, since more of the blue light is absorbed, objects seen through a dust layer appear redder. 
The general extinction law to deredden IR to UV spectrophotometric data is given by:
\begin{equation}
\begin{aligned}
\frac{A_{V}}{E(B-V)}=R_{V},
\end{aligned}
\label{eq:extinction1}
\end{equation}
where $A_{V}$ is the extinction in optical $V$ band and $E(B-V)$ is the colour-reddening. According to \citet{cardelli1989} this ratio has an average value of $R_{V}=3.1$ toward the Galactic diffuse interstellar medium (ISM) line-of-sight. 

The value of colour-reddening can be derived from the \citet[][]{Schlegel1998} dust maps in which they presented a full-sky 100 $\mu$m map that can be converted to a map proportional to dust column density. A detailed follow-up analysis of dust reddening was carried out by \citet{schlafly2011} using the colours of stars in the Sloan Digital Sky Survey (SDSS). It was found that \citet{Schlegel1998} overestimated reddening by about 14\% in ($B-V$) and this is therefore seen in all colours. Consequently a factor of 0.86 must be applied to the \citet{Schlegel1998}  extinction maps for recalibration. This overestimation was proven to be even higher at low Galactic latitudes, $|b|<5\deg$. For example \citet{nagayama2004} determined  the total foreground extinction by  measuring the colour excess for giant stars around the radio galaxy PKS 1343-601 and found the extinction to be overestimated by 33\% when extrapolating Schlegel maps to their survey region. Numerous other studies such as \citet{arce1999,dutra2002,vandriel2009,schroder2007} confirm the overestimate of $f=0.86$ for latitudes of $|b|<5\deg$. In this work we adopt the \citet{Schlegel1998} $E(B-V)$ values corrected by a factor of 0.86.

Using the \citet{cardelli1989} parametrisation, the extinctions in $J$, $H$ and $K_{s}$-bands are: 

\begin{equation}
\begin{aligned}
A_{J}&=0.863E(B-V),\\
A_{H}&=0.570E(B-V),\\
A_{K_{s}}&=0.368E(B-V).\\
\end{aligned}
\label{eq:extinction2}
\end{equation}
 
Equation \ref{eq:extinction2} is employed to derive the extinction-corrected magnitudes. The mean, minimum and maximum values of extinction derived at the position of the detected galaxies in the clusters of the VSCL region are listed in Table~\ref{tab:foregroundext}. 

\begin{table}
\small
\caption{The mean, minimum and maximum values of $A_{J},A_{H}$ and $A_{K_{s}}$ for the galaxies within the VSCL clusters.}
\begin{center}
\scalebox{0.95}{
\begin{tabular}{ccccc}
\toprule
\toprule
  & & $A_{J}$ & $A_{H}$ & $A_{K_s}$\\
   & & (mag)&(mag)&(mag)\\
\midrule

VC02 & mean & $0.249 \pm 0.024$ & $0.164\pm 0.016$ & $0.106\pm 0.010$ \\
     &min & $0.191$ & $0.126$ & $0.081$  \\
     &max & $0.310$ & $0.204$ & $0.132$  \\     
\midrule
VC04 & mean  & $0.166 \pm 0.015$ & $0.110 \pm 0.010$ & $0.071 \pm 0.006$ \\
     & min & $0.147$ & $0.097$ & $0.062$  \\
     & max & $0.204$ & $0.135$ & $0.087$  \\     
\midrule
VC05 & mean & $0.147 \pm 0.011$ & $0.097\pm 0.007$ & $0.063\pm 0.005$\\
     & min & $0.125$ & $0.083$ & $0.053$\\
     & max & $0.185$ & $0.122$ & $0.079$\\     
\midrule
VC08 & mean & $0.197\pm 0.014$ & $0.130\pm 0.009$ & $0.084\pm 0.006$ \\
     & min & $0.180$ & $0.119$ & $0.076$ \\
     & max & $0.226$ & $0.149$ & $0.096$ \\     
\midrule
VC10 & mean & $0.188\pm 0.013$ & $0.124\pm 0.008$ & $0.080\pm 0.005$  \\
     & min & $0.170$ & $0.112$ & $0.072$\\
     & max & $0.235$ & $0.155$ & $0.100$\\     
\midrule
VC11 & mean & $0.201\pm 0.041$ & $0.132\pm 0.027$ & $0.086\pm 0.018$ \\
     & min & $0.141$ & $0.093$ & $0.060$ \\
     & max & $0.302$ & $0.200$ & $0.129$ \\     
\bottomrule
\end{tabular}}
\end{center}
\label{tab:foregroundext}
\end{table}

\begin{figure*}
\centering
\begin{subfigure}[b]{0.44\textwidth}\includegraphics[width=\columnwidth, height=\textheight,keepaspectratio]{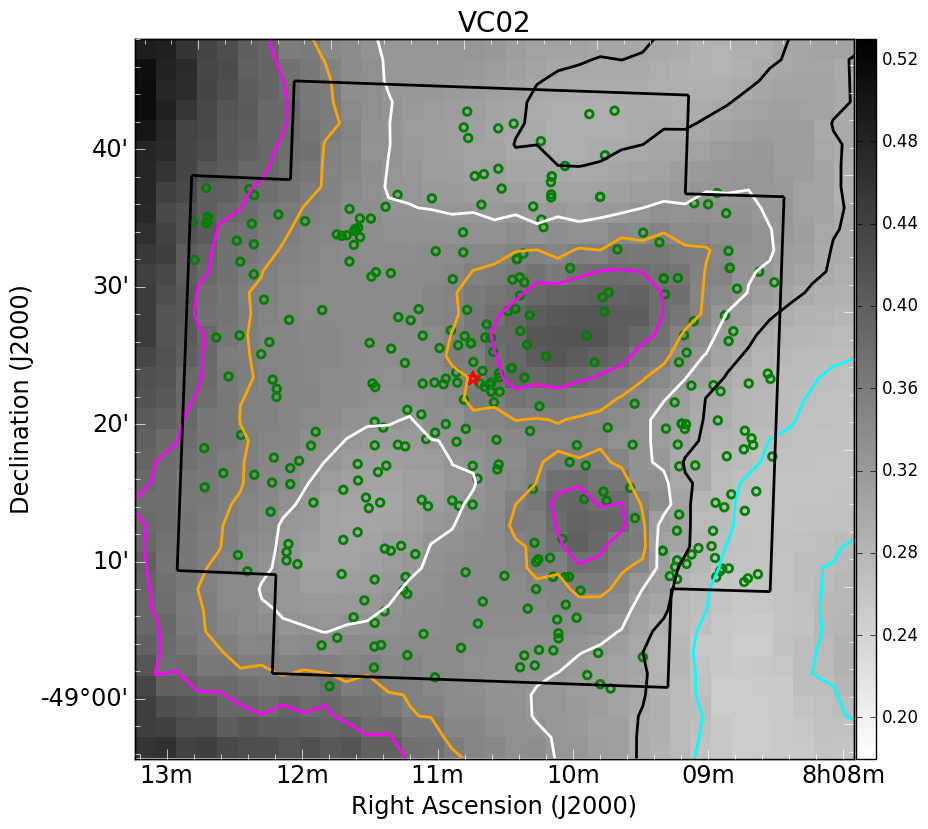}\end{subfigure}
\begin{subfigure}[b]{0.44\textwidth}\includegraphics[width=\columnwidth, height=\textheight, keepaspectratio]{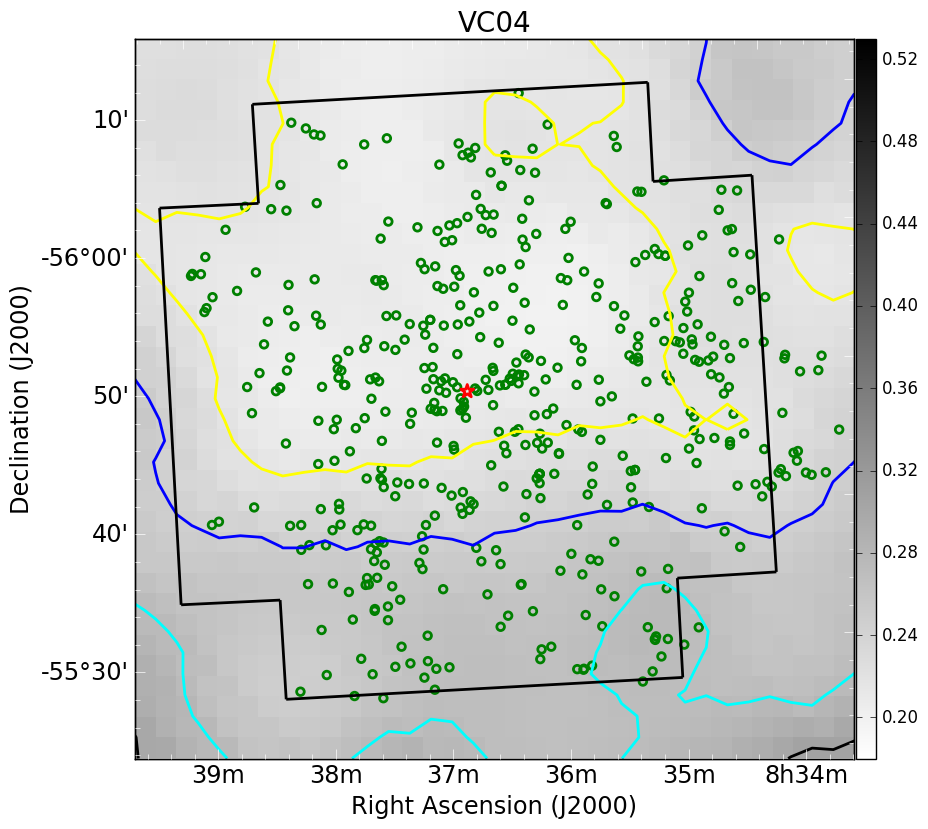}\end{subfigure}

\begin{subfigure}[b]{0.44\textwidth}\includegraphics[width=\columnwidth, height=\textheight, keepaspectratio]{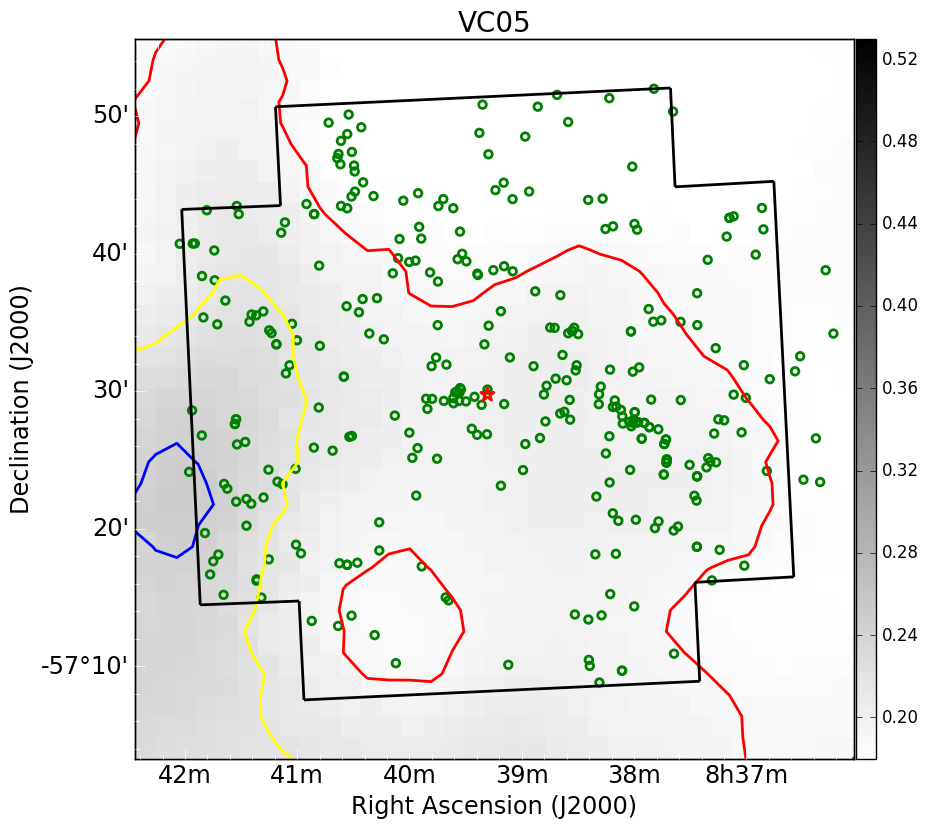}\end{subfigure}
\begin{subfigure}[b]{0.44\textwidth}\includegraphics[width=\columnwidth, height=\textheight, keepaspectratio]{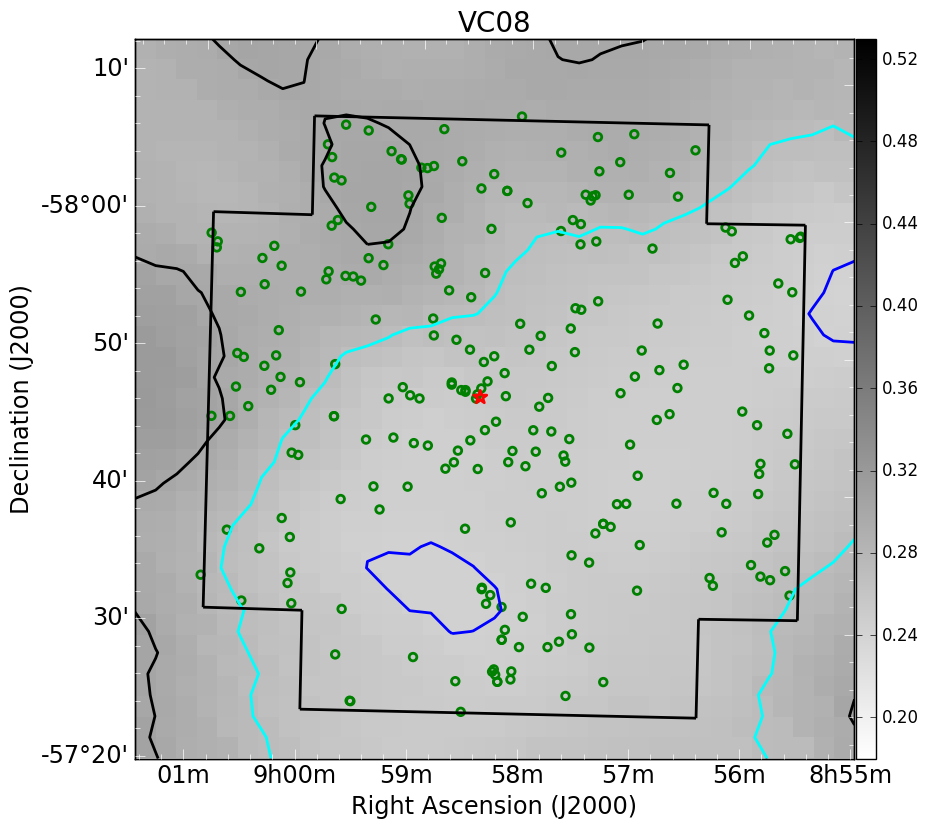}\end{subfigure}

\begin{subfigure}[b]{0.44\textwidth}\includegraphics[width=\columnwidth, height=\textheight, keepaspectratio]{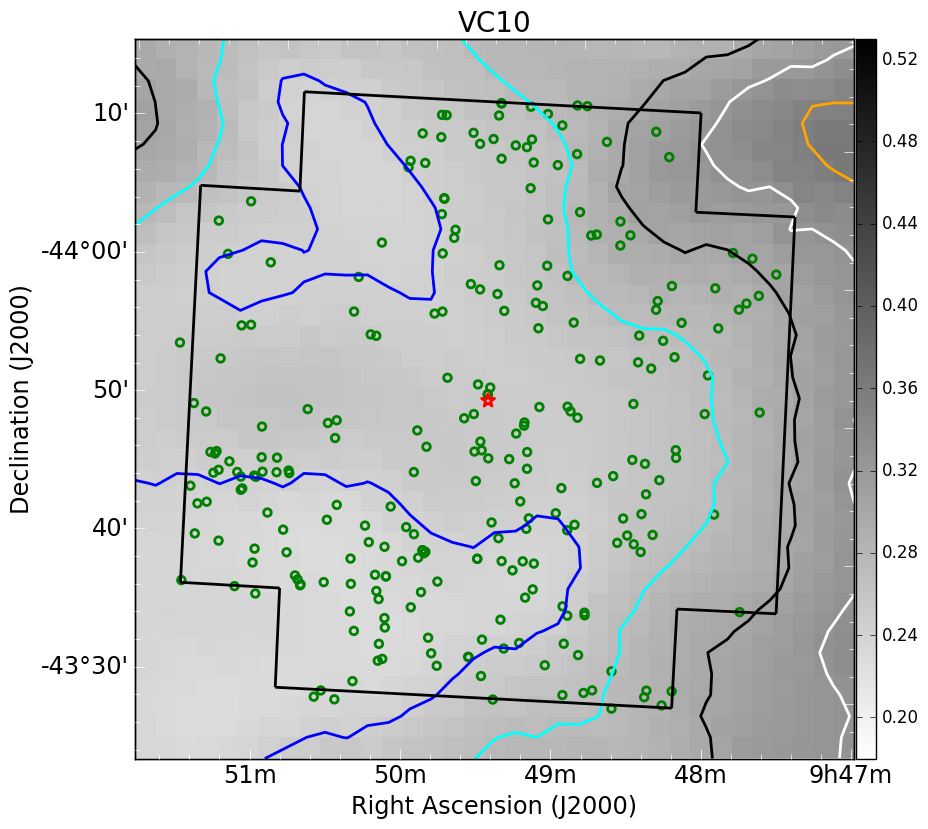}\end{subfigure}
\begin{subfigure}[b]{0.44\textwidth}\includegraphics[width=\columnwidth, height=\textheight, keepaspectratio]{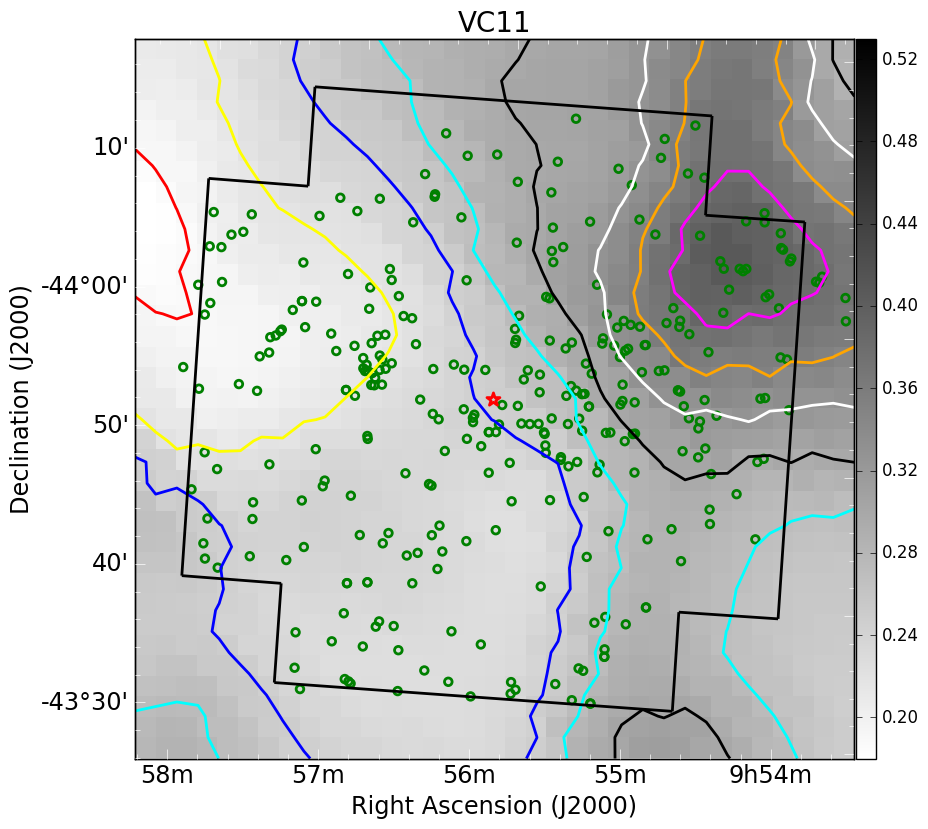}\end{subfigure}

\caption{The reddening maps of the region around the VSCL clusters. The green circles are the galaxy candidates in the galaxy clusters and the contours show the $K_s$-band extinction levels, with red indicating $A_{Ks}=0\fm07$, yellow $A_{Ks}=0\fm08$, blue $A_{Ks}=0\fm09$, cyan $A_{Ks}=0\fm10$, black $A_{Ks}=0\fm11$, white $A_{Ks}=0\fm12$, orange $A_{Ks}=0\fm13$, magenta $A_{Ks}=0\fm14$. The black boxes indicate the outlines of the observed mosaic in each cluster. The galaxies outside of the boxes are located in the extra fields that were observed to compensate for the Infrared detector problem of the IRSF telescope. The colourbar shows the reddening map values as grey-scale.}
\label{fig:extinction}
\end{figure*}

In Fig.~\ref{fig:extinction} the galaxy distributions of the six analysed clusters are plotted on the extracts of the \citet{Schlegel1998} extinction maps\footnote{Downloaded from \url{https://irsa.ipac.caltech.edu/applications/DUST/}}. The green open circles represent the positions of the candidate galaxies and the contours show the extinction levels in the $K_s$-band ($A_{K_{s}}$). These figures indicate that amongst these six clusters VC02 is located in the highest extinction area ($\langle A_{K_{s}} \rangle = 0.106\pm 0.010$),  and VC04 and VC05 have the lowest foreground extinction ($\langle A_{K_{s}} \rangle = 0.071\pm 0.006$ and $0.063\pm 0.005$ respectively). The foreground extinction in the regions of VC08, VC10 and VC11 is roughly in the same range ($\langle A_{K_{s}} \rangle = 0.084\pm 0.006$, $0.080\pm 0.005$ and $0.086\pm 0.018$ respectively). Also there are more patches of higher extinction in VC02 than the other clusters. The final analysis is based on the extinction-corrected magnitudes.

\subsection{Extinction effect on the properties of galaxies}

In addition to foreground extinction, Galactic extinction results in galaxies appearing smaller than they really are. The low surface brightness outer parts  disappear as the extinction increases and results in a smaller apparent radii.
The extinction will also affect the derivation of total magnitudes  since the fainter outer parts of a galaxy cannot be fully retrieved with the extrapolation of the surface brightness profiles \citep{cameron1990,nagayama2004,riadpaper2010,said2014}. 

The isophotal magnitudes are preferable to total magnitudes when working with ZoA surveys. This was  shown in 
\citet{said2014} who compared the total and isophotal magnitudes of a sample of 2MASX ZoA galaxies with much deeper IRSF photometry. The observational set-up was the same as this survey, i.e. $\sim 2\fm0$ deeper than 2MASX.  Figure~1 in \citet{said2014} indicates that the mean offset between 2MASX and IRSF for the total magnitudes is considerable while the offset is insignificant for the isophotal magnitudes. We therefore, only use the $K_{s20}$ fiducial isophotal magnitudes for the VSCL cluster analyses in this survey.

We applied the empirical relations derived by \citet{riadpaper2010} for early-type galaxies to correct for extinction effects on our $J$, $H$, $K_s$  isophotal magnitudes.
We implemented this method to assess the additional $\mu_c$-optimized extinction effect on the isophotal magnitudes of early-type galaxies in the VSCL clusters (early-type galaxies are dominant in galaxy clusters \citep[morphology-density relation, ][]{dressler1980}). The results are shown in Table~\ref{tab:additionext-E}. It reveals that the effect of the $\mu_c$-optimized extinction on the magnitudes is significantly smaller than the overall foreground extinction ($A_\lambda$, see Table~\ref{tab:foregroundext}), and much smaller than the uncertainties of the isophotal magnitudes of galaxies (even for VC02 which suffers from the highest foreground extinction and largest variation over the mosaic). We therefore concluded that it is sufficient to apply the \citet{schlafly2011} foreground values to the isophotal magnitudes without any further correction for loss of the outer LSB features.

\begin{table}
\caption{The mean values of the extinction effect corrections on the $J$, $H$, $K_s$ isophotal magnitudes of galaxies distributed over the regions of the six VSCL clusters (i.e. $\Delta m_{iso}$), derived from the empirical relations in \citet{riadpaper2010} for early-type galaxies.} 

\begin{center}
\scalebox{0.95}{
\begin{tabular}{cccc}
\toprule
\toprule
Cluster & $\langle \Delta J_{iso} \rangle$ & $\langle \Delta H_{iso} \rangle$ & $ \langle \Delta K_{iso} \rangle$\\
    & (mag)&(mag)&(mag)\\
\midrule
VC02  &0.040 & 0.048 & 0.006 \\
VC04 & 0.021 & 0.028 & 0.003\\
VC05  & 0.018 & 0.025 & 0.002 \\
VC08  & 0.028 & 0.036 & 0.004 \\
VC10  & 0.027 & 0.036 & 0.003 \\
VC11  & 0.028 & 0.036 & 0.004 \\

\bottomrule
\end{tabular}}
\end{center}
\label{tab:additionext-E}
\end{table}

\subsection{Extinction-Corrected NIR colours}
Using equation \ref{eq:extinction2}, the extinction-corrected NIR colours are derived as a function of $E(B-V)$:
\begin{equation}
\begin{aligned}
(J-H)^{o}&=(J-H)-0.293 E(B-V),\\
(H-K_{s})^{o}&=(H-K_{s})-0.202 E(B-V),\\
(J-K_{s})^{o}&=(J-K_{s})-0.495 E(B-V),\\
\end{aligned}
\label{eq:nircolours}
\end{equation}

Colours can be used as a tool in NIR to distinguish galaxies from stars. Galaxies appear red on $J$, $H$, $K_s$ composite images due to the old (red) population of stars in their bulges or cores. In addition, the extinction effect in the ZoA causes galaxies to look even redder.

\begin{figure*}
\centering
  \includegraphics[width=1.8\columnwidth, height=\textheight, keepaspectratio]{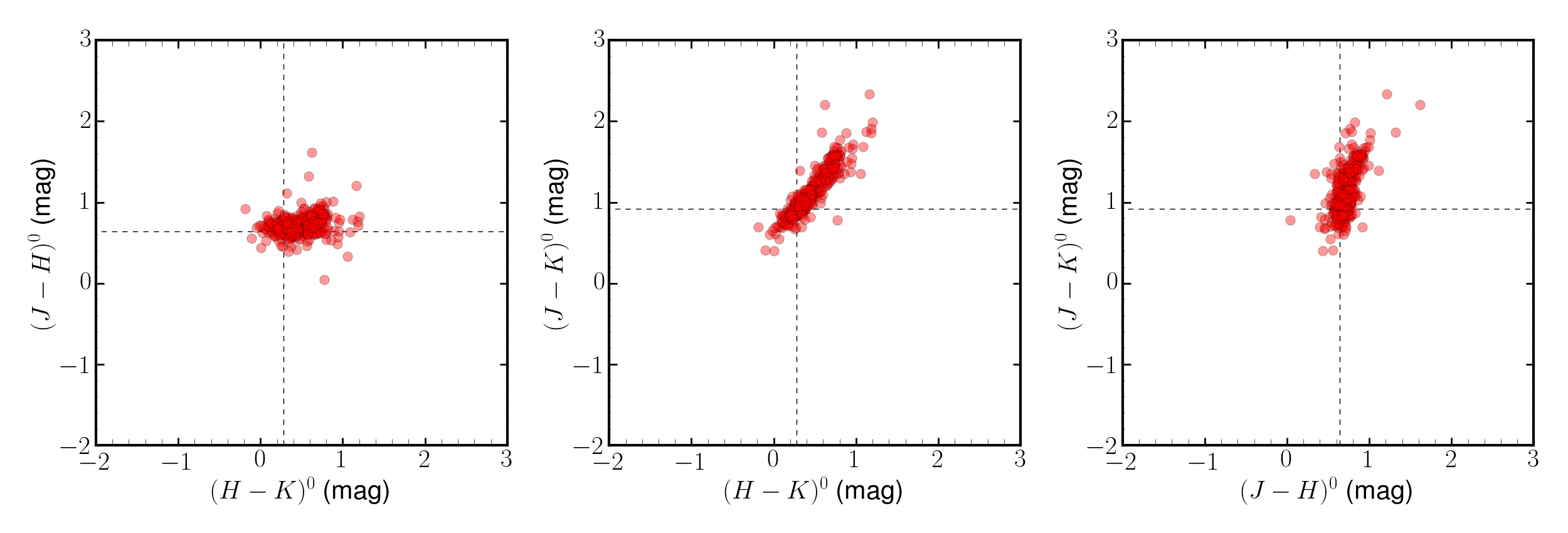}
  \caption{Extinction-corrected colour-colour diagrams of galaxy candidates with $r>2''$ in VC04. The dashed lines indicate the mean NIR isophotal colours for nearby early-type galaxies \citep{jarrett2000b-k}.}
  \label{fig:c-c-vc04}
\end{figure*}

 Unlike the optical colours, the NIR colours hardly  depend on the morphology of the galaxies \citep{jarrett2000b-k}. However, the late spirals (e.g. Sd/Sdm) are bluer compared to other spiral and elliptical galaxies. Therefore, we also used extinction-corrected colour-colour diagrams to distinguish between galaxies and extended Galactic objects as shown in Fig.~\ref{fig:c-c-vc04}.  \citet{jarrett2000b-k} determined the mean NIR isophotal colours as $(J-H)$=0\fm64, $(H-K_{s})$=0\fm28 and $(J-K_{s})$=0\fm92 for nearby (low redshift) early-type and $(J-K_{s})=$0\fm99 for late-type spiral galaxies. In colour-colour diagrams the candidates far from the intersection of mean NIR colours are unlikely to be galaxies. 
 
However, at higher redshifts the emission of galaxies shifts from the $H$ to $K_{s}$-band and as a result ($H-K_{s}$) becomes redder while ($J-H$) remains roughly constant \citep[cosmic reddening,][]{jarrett2004}. This effect is particularly obvious in the $(J-K_{s})^o$ vs $(H-K_{s})^o$ diagram (see the middle panel of Fig.~\ref{fig:c-c-vc04}), where the reddest points in ($H-K_{s}$) and ($J-K_{s}$) might actually be AGNs which mostly are intrinsically red (because of dust), and distant (hence have a high $k$-correction).

\subsection{$k$-Correction in the VSCL clusters}
\label{sec:kcorrection}
The expansion of the universe causes the rest-frame wavelength of an object to shift to longer wavelengths as a function of redshift. The apparent magnitude of any astronomical object needs to be converted to an equivalent measurement of its rest-frame by the so-called $k$-correction. 

\begin{figure*}
    \includegraphics[width=1.8\columnwidth, height=\textheight, keepaspectratio]{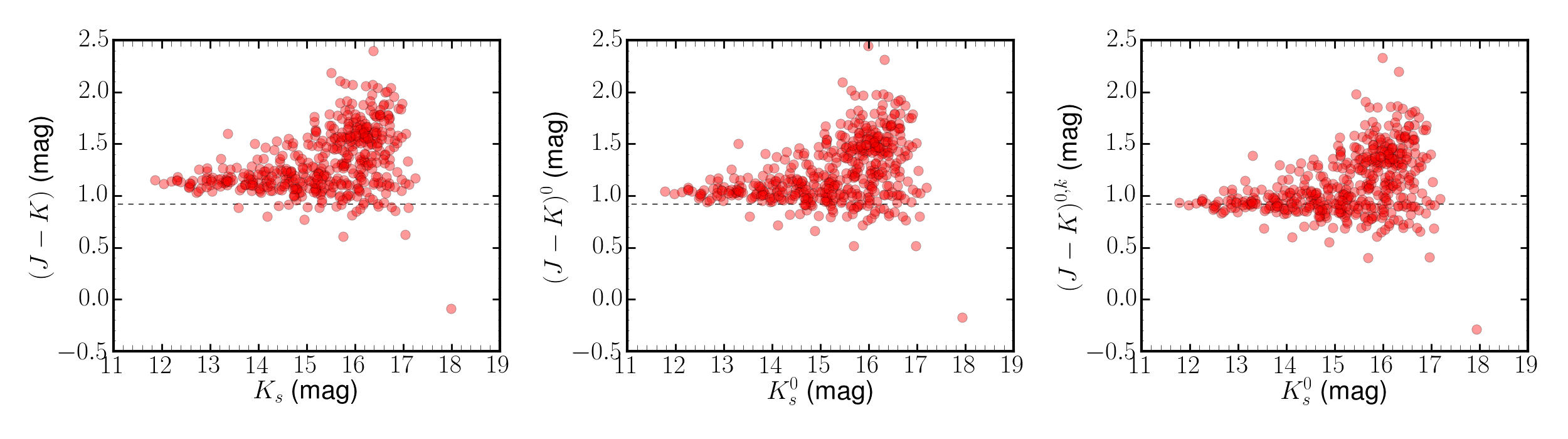}
    
\caption{The colour-magnitude diagram ($(J-K_{s})$ versus $K_{s}$) for galaxy candidates in VC04. From left to right the plots are observed, extinction-corrected and extinction plus $k$-corrected. The dashed lines indicate the mean NIR isophotal colour $(J-K_{s})$=0\fm92 for nearby early-type galaxies \citep{jarrett2000b-k}.}  
\label{fig:kcor-c-m}
\end{figure*}

We implemented the $k$-correction derived by T.~Jarrett (private communication) on NIR colours of galaxies in the VSCL clusters. At the redshift of the VSCL, $cz= 18000\,$km~s\textsuperscript{-1} KK2017, the $k$-correction for $(J-K_{s})$ is $\sim 0\fm112$ for early-type galaxies. As an example, we display this effect for VC04 in Fig.~\ref{fig:kcor-c-m}, which has the largest number of identified galaxy candidates. Assuming that all galaxy candidates lie at the VSCL distance, we plot in Fig.~\ref{fig:kcor-c-m} the colour-magnitude diagram for VC04 potential galaxies before and after foreground extinction-correction (left and middle panels respectively) and including the $k$-correction as well (right panel). The plot clearly reveals a redshift offset which, although small, is not insignificant ($\Delta m = 0\fm112$). After the $k$-correction the mean $(J-K_s)$ colour is closer to the mean colour of nearby early-type galaxies, $(J-K_{s})=0\fm92$ \citep{jarrett2000b-k}.  
It is clear from Fig.~\ref{fig:kcor-c-m} that the majority of the cluster members lie around the expected mean colour value, however, at the faint end there remains an increasing number of highly reddened galaxies, which are not cluster members, but rather background galaxies.

\section{NIR Extended source catalogue}
\label{sec:nircatalog}
\subsection{Catalogue parameters}

\label{subsec:catpar}
The output catalogue from the IRSF pipeline contains 125 parameters for each galaxy which are listed in \citet{williams2011}. Six catalogues were prepared containing the astrometric and photometric parameters of 1715 identified galaxies distributed over six clusters, of which only $\sim 15\%$ were previously known. There are 290, 416, 279, 217, 232 and 281 identified galaxies with $r>2''$ in VC02, VC04, VC05, VC08, VC10 and VC11 respectively (for details see Table~\ref{tab:5clustersprop}). Here we describe the NIR parameters that are the most relevant for the subsequent analysis. As an example we present a sample output page of the catalogue for the cluster VC08 in Table~\ref{tab:irsfcat}. The full catalogues of galaxies for all of the six VSCL cluster candidates are available online. The dashes in Table~\ref{tab:irsfcat} refer to the total magnitude values which could not be measured by the pipeline.\\

\noindent {\sl Column }$1$: Unique identifier; (designation)

\noindent [ZOAhhmmss.sss$\pm$ddmmss.ss].

\noindent {\sl Column }$2, 3$: Galactic coordinates; ($\ell$, $b$) [deg].

\noindent {\sl Column }$4$: $J$-band ellipticity ($\epsilon = 1 -b/a$).

\noindent {\sl Column }$5$: $J$-band position angle (East of North); ($\phi_J$) [deg].

\noindent {\sl Column }$6$: $K_{s20}$ fiducial isophotal radius; ($r_{K20fe}$) [arcsec].

\noindent {\sl Column }$7$: $J$-band $K_{s20}$ fiducial isophotal magnitude and error; ($J_{K20fe}$) [mag].

\noindent {\sl Column }$8$: $H$-band $K_{s20}$ fiducial isophotal magnitude and error; ($H_{K20fe}$) [mag].

\noindent {\sl Column }$9$: $K_s$-band $K_{s20}$ fiducial isophotal magnitude and error; ($K_{K20fe}$) [mag].

\noindent {\sl Column }$10$: $J$-band extrapolated total magnitude and error; ($J_{tot}$) [mag].

\noindent {\sl Column }$11$: $H$-band extrapolated total magnitude and error; ($H_{tot}$) [mag].

\noindent {\sl Column }$12$: $K_s$-band extrapolated total magnitude and error; ($K_{stot}$) [mag].

\noindent {\sl Column }$13-15$: $J$, $H$ and $K_s$ central surface brightness; ($\mu_{J}$, $\mu_{H}$, $\mu_{Ks}$) [mag arcsec\textsuperscript{-2}], defined as $\mu_\lambda=m_\lambda (3'') + log (3.0^2 \times \pi)$.

\noindent {\sl Column }$16$: Galactic reddening along the line of sight \citep{Schlegel1998}; ($E(B-V)$) [mag].

\begin{sidewaystable*}
\scriptsize
\caption{The VC08 IRSF pipeline output sample catalogue. The dashes indicate that the pipeline could not fit a total magnitude. The full table is available online.}

\begin{tabular}{ccccrrcccccccccc}
\toprule
\toprule
1&2&3&4&5&6&7&8&9&10&11&12&13&14&15&16\\
\midrule
Designation & $\ell$ & $b$ & $e_J$ & $\phi_J$ & $r_{k20fe}$ & $J_{k20fe}$ & $H_{k20fe}$ & $K_{s_{k20fe}}$ & $J_{tot}$ & $H_{tot}$ & $K_{s_{tot}}$ & $\mu_{J}$ &$\mu_{H}$ &$\mu_{Ks}$ &$E(B-V)$  \\
 & (deg) &(deg) & &(deg) & (arcsec) & (mag) & (mag) & (mag) & (mag) & (mag) & (mag)  & & (mag arcsec\textsuperscript{-2})  & & (mag) \\
\midrule 
$ZOA085648.318-575226.01$ & $275.628$ & $-8.047$ & $0.14$ & $50.74$ & $2.84$ & $17.13\pm 0.07$ & $16.66\pm 0.08$ & $15.87\pm 0.10$ & $16.58\pm 0.20$ & $15.74\pm 0.29$ & $15.80\pm 0.10$ & $20.68$ & $20.03$ & $19.02$ & $0.218$ \\
$ZOA085707.906-574718.60$ & $275.590$ & $-7.959$ & $0.23$ & $28.81$ & $6.35$ & $15.35\pm 0.03$ & $14.71\pm 0.03$ & $14.23\pm 0.04$ & $-$ & $14.34\pm 0.07$ & $13.35\pm 0.18$ & $18.77$ & $18.11$ & $17.63$ & $0.218$ \\
$ZOA085636.836-574745.78$ & $275.552$ & $-8.017$ & $0.29$ & $-45.14$ & $11.33$ & $13.45\pm 0.02$ & $12.69\pm 0.02$ & $12.31\pm 0.03$ & $13.18\pm 0.11$ & $-$ & $11.57\pm 0.20$ & $17.50$ & $16.76$ & $16.38$ & $0.218$ \\
$ZOA085700.212-574832.51$ & $275.595$ & $-7.985$ & $0.11$ & $-39.62$ & $2.88$ & $17.51\pm 0.09$ & $16.51\pm 0.07$ & $15.59\pm 0.07$ & $-$ & $-$ & $14.74\pm 0.25$ & $20.34$ & $19.47$ & $18.91$ & $0.218$ \\
$ZOA085646.873-574902.59$ & $275.583$ & $-8.013$ & $0.30$ & $82.70$ & $4.22$ & $16.55\pm 0.06$ & $15.75\pm 0.05$ & $15.33\pm 0.07$ & $15.60\pm 0.40$ & $14.53\pm 0.36$ & $-$ & $19.57$ & $18.96$ & $18.47$ & $0.218$ \\
$ZOA085656.746-575026.96$ & $275.615$ & $-8.012$ & $0.16$ & $40.71$ & $6.71$ & $14.34\pm 0.03$ & $13.60\pm 0.02$ & $13.22\pm 0.03$ & $13.61\pm 0.24$ & $13.20\pm 0.09$ & $12.91\pm 0.06$ & $17.95$ & $17.21$ & $16.83$ & $0.218$ \\
$ZOA085640.877-574550.96$ & $275.533$ & $-7.990$ & $0.12$ & $57.58$ & $3.67$ & $15.33\pm 0.03$ & $14.55\pm 0.03$ & $14.00\pm 0.03$ & $15.02\pm 0.09$ & $13.24\pm 0.39$ & $11.86\pm 0.68$ & $18.63$ & $17.90$ & $17.36$ & $0.218$ \\
$ZOA085647.800-574524.99$ & $275.537$ & $-7.973$ & $0.21$ & $-4.19$ & $4.84$ & $15.33\pm 0.04$ & $14.56\pm 0.03$ & $14.19\pm 0.04$ & $-$ & $14.03\pm 0.12$ & $-$ & $18.69$ & $17.92$ & $17.55$ & $0.218$ \\
$ZOA085633.644-574928.12$ & $275.569$ & $-8.041$ & $0.10$ & $-85.66$ & $3.89$ & $17.04\pm 0.08$ & $16.14\pm 0.06$ & $15.27\pm 0.07$ & $13.39\pm 1.28$ & $14.46\pm 0.55$ & $-$ & $20.16$ & $19.30$ & $18.50$ & $0.218$ \\
$ZOA085651.875-575753.11$ & $275.704$ & $-8.099$ & $0.38$ & $31.61$ & $3.38$ & $16.59\pm 0.05$ & $15.95\pm 0.04$ & $15.83\pm 0.08$ & $-$ & $15.68\pm 0.04$ & $-$ & $19.64$ & $19.09$ & $19.20$ & $0.235$ \\
$ZOA085638.371-580142.02$ & $275.734$ & $-8.163$ & $0.05$ & $-32.06$ & $2.92$ & $16.19\pm 0.04$ & $15.46\pm 0.03$ & $14.84\pm 0.04$ & $-$ & $14.59\pm 0.27$ & $14.43\pm 0.09$ & $19.38$ & $18.75$ & $18.10$ & $0.238$ \\
$ZOA085643.022-580324.85$ & $275.763$ & $-8.173$ & $0.35$ & $-1.65$ & $2.10$ & $17.50\pm 0.08$ & $16.60\pm 0.06$ & $16.13\pm 0.10$ & $-$ & $-$ & $-$ & $20.38$ & $19.70$ & $19.56$ & $0.238$ \\
$ZOA085710.532-580408.06$ & $275.811$ & $-8.135$ & $0.34$ & $48.53$ & $9.97$ & $13.27\pm 0.02$ & $12.49\pm 0.02$ & $12.13\pm 0.02$ & $-$ & $11.94\pm 0.16$ & $12.00\pm 0.03$ & $16.98$ & $16.20$ & $15.82$ & $0.238$ \\
$ZOA085629.176-580505.56$ & $275.765$ & $-8.215$ & $0.08$ & $-70.08$ & $2.92$ & $17.21\pm 0.07$ & $16.38\pm 0.06$ & $15.66\pm 0.08$ & $-$ & $-$ & $15.34\pm 0.09$ & $20.18$ & $19.47$ & $18.81$ & $0.238$ \\
$ZOA085703.083-580611.25$ & $275.827$ & $-8.169$ & $0.26$ & $11.66$ & $2.01$ & $18.09\pm 0.13$ & $17.17\pm 0.09$ & $16.11\pm 0.10$ & $-$ & $-$ & $15.90\pm 0.11$ & $21.39$ & $20.31$ & $20.42$ & $0.238$ \\
$ZOA085705.491-580146.31$ & $275.774$ & $-8.118$ & $0.09$ & $76.54$ & $2.51$ & $17.79\pm 0.10$ & $16.67\pm 0.06$ & $16.12\pm 0.09$ & $-$ & $-$ & $-$ & $20.55$ & $19.85$ & $19.52$ & $0.238$ \\
$ZOA085834.283-572605.83$ & $275.443$ & $-7.584$ & $0.22$ & $50.20$ & $3.07$ & $16.72\pm 0.05$ & $15.83\pm 0.04$ & $15.30\pm 0.06$ & $-$ & $14.69\pm 0.33$ & $14.96\pm 0.08$ & $19.82$ & $19.02$ & $18.49$ & $0.215$ \\
$ZOA085811.545-572607.43$ & $275.410$ & $-7.623$ & $0.13$ & $46.94$ & $4.06$ & $16.07\pm 0.05$ & $15.26\pm 0.04$ & $14.62\pm 0.05$ & $-$ & $-$ & $13.94\pm 0.15$ & $19.26$ & $18.55$ & $17.90$ & $0.215$ \\
$ZOA085812.792-572637.97$ & $275.418$ & $-7.627$ & $0.28$ & $-75.44$ & $4.67$ & $15.52\pm 0.04$ & $14.71\pm 0.03$ & $14.42\pm 0.05$ & $15.37\pm 0.07$ & $-$ & $13.28\pm 0.31$ & $18.91$ & $18.12$ & $17.77$ & $0.215$ \\
$ZOA085830.956-572352.63$ & $275.409$ & $-7.566$ & $0.17$ & $-64.48$ & $3.72$ & $17.32\pm 0.10$ & $16.37\pm 0.07$ & $15.44\pm 0.08$ & $-$ & $-$ & $-$ & $20.47$ & $19.49$ & $18.79$ & $0.215$ \\
$ZOA085857.503-572746.54$ & $275.498$ & $-7.562$ & $0.06$ & $-66.49$ & $4.11$ & $16.37\pm 0.05$ & $15.63\pm 0.04$ & $15.49\pm 0.09$ & $14.94\pm 0.39$ & $14.62\pm 0.15$ & $13.01\pm 0.62$ & $19.36$ & $18.63$ & $18.55$ & $0.215$ \\
$ZOA085814.502-572651.08$ & $275.424$ & $-7.626$ & $0.68$ & $77.28$ & $2.59$ & $17.83\pm 0.11$ & $16.77\pm 0.07$ & $16.05\pm 0.09$ & $-$ & $-$ & $12.58\pm 0.98$ & $21.20$ & $19.72$ & $19.03$ & $0.215$ \\
$ZOA085813.602-572701.32$ & $275.424$ & $-7.629$ & $0.25$ & $24.06$ & $2.25$ & $18.06\pm 0.12$ & $17.04\pm 0.08$ & $16.31\pm 0.11$ & $-$ & $-$ & $16.12\pm 0.12$ & $20.84$ & $20.07$ & $19.61$ & $0.215$ \\
$ZOA085809.682-572911.78$ & $275.447$ & $-7.659$ & $0.12$ & $36.69$ & $3.42$ & $17.05\pm 0.10$ & $16.38\pm 0.08$ & $15.66\pm 0.11$ & $16.28\pm 0.32$ & $-$ & $11.31\pm 1.20$ & $19.86$ & $19.34$ & $18.57$ & $0.215$ \\
$ZOA085821.261-573250.20$ & $275.510$ & $-7.679$ & $0.21$ & $-8.35$ & $2.43$ & $17.06\pm 0.06$ & $16.16\pm 0.04$ & $15.65\pm 0.07$ & $11.16\pm 1.90$ & $-$ & $14.14\pm 0.34$ & $20.05$ & $19.18$ & $18.61$ & $0.209$ \\
$ZOA085831.042-573712.41$ & $275.581$ & $-7.709$ & $0.22$ & $-31.29$ & $2.97$ & $16.72\pm 0.05$ & $16.03\pm 0.04$ & $15.37\pm 0.06$ & $11.48\pm 8.94$ & $-$ & $-$ & $19.81$ & $19.22$ & $18.46$ & $0.209$ \\
$ZOA085816.493-573225.22$ & $275.498$ & $-7.682$ & $0.67$ & $-0.16$ & $7.30$ & $15.51\pm 0.04$ & $14.66\pm 0.03$ & $14.19\pm 0.04$ & $15.15\pm 0.11$ & $13.82\pm 0.24$ & $13.16\pm 0.31$ & $18.96$ & $18.14$ & $17.66$ & $0.209$ \\
$ZOA085818.652-573146.87$ & $275.493$ & $-7.672$ & $0.19$ & $45.90$ & $2.90$ & $16.68\pm 0.06$ & $15.85\pm 0.04$ & $15.23\pm 0.06$ & $-$ & $15.12\pm 0.21$ & $14.49\pm 0.20$ & $19.97$ & $19.07$ & $18.41$ & $0.209$ \\
$ZOA085821.087-573256.46$ & $275.511$ & $-7.680$ & $0.05$ & $-49.88$ & $4.25$ & $17.71\pm 0.10$ & $17.01\pm 0.08$ & $16.23\pm 0.12$ & $-$ & $-$ & $15.49\pm 0.19$ & $20.49$ & $19.79$ & $19.18$ & $0.209$ \\
$ZOA085815.583-574502.92$ & $275.659$ & $-7.820$ & $0.45$ & $33.42$ & $6.82$ & $14.49\pm 0.03$ & $13.79\pm 0.02$ & $13.38\pm 0.03$ & $12.97\pm 0.47$ & $12.42\pm 0.38$ & $12.97\pm 0.04$ & $17.96$ & $17.26$ & $16.87$ & $0.216$ \\
$ZOA085902.891-574009.90$ & $275.665$ & $-7.687$ & $0.25$ & $53.21$ & $3.56$ & $15.89\pm 0.04$ & $15.14\pm 0.03$ & $14.80\pm 0.05$ & $-$ & $13.31\pm 0.52$ & $11.01\pm 1.07$ & $19.21$ & $18.38$ & $17.99$ & $0.216$ \\
$ZOA085824.933-574134.48$ & $275.628$ & $-7.766$ & $0.52$ & $-72.60$ & $5.50$ & $15.10\pm 0.03$ & $14.40\pm 0.03$ & $13.94\pm 0.03$ & $14.74\pm 0.13$ & $13.86\pm 0.16$ & $13.73\pm 0.04$ & $18.56$ & $17.87$ & $17.41$ & $0.216$ \\
$ZOA085842.568-574132.50$ & $275.653$ & $-7.736$ & $0.22$ & $-66.53$ & $8.02$ & $13.90\pm 0.02$ & $13.15\pm 0.02$ & $12.81\pm 0.03$ & $13.24\pm 0.22$ & $-$ & $12.43\pm 0.10$ & $17.57$ & $16.82$ & $16.46$ & $0.216$ \\
$ZOA085838.007-574201.32$ & $275.653$ & $-7.749$ & $0.08$ & $-17.91$ & $2.65$ & $15.63\pm 0.03$ & $15.01\pm 0.03$ & $14.70\pm 0.04$ & $-$ & $14.78\pm 0.05$ & $-$ & $19.01$ & $18.35$ & $18.26$ & $0.216$ \\
$ZOA085836.000-574252.69$ & $275.661$ & $-7.762$ & $0.22$ & $19.23$ & $3.03$ & $16.78\pm 0.05$ & $15.90\pm 0.04$ & $15.16\pm 0.05$ & $-$ & $-$ & $14.83\pm 0.09$ & $19.95$ & $19.12$ & $18.39$ & $0.216$ \\
$ZOA085852.452-574311.59$ & $275.689$ & $-7.737$ & $0.38$ & $33.74$ & $4.04$ & $16.64\pm 0.05$ & $16.02\pm 0.04$ & $15.26\pm 0.06$ & $-$ & $15.79\pm 0.08$ & $14.98\pm 0.08$ & $20.17$ & $19.43$ & $18.77$ & $0.216$ \\
$ZOA085900.120-574320.52$ & $275.702$ & $-7.726$ & $0.40$ & $78.68$ & $3.12$ & $16.55\pm 0.05$ & $15.81\pm 0.04$ & $15.35\pm 0.06$ & $16.20\pm 0.10$ & $-$ & $15.13\pm 0.08$ & $19.70$ & $19.03$ & $18.62$ & $0.216$ \\
$ZOA085821.509-574424.85$ & $275.659$ & $-7.803$ & $0.47$ & $06.72$ & $7.13$ & $15.02\pm 0.03$ & $14.27\pm 0.03$ & $13.90\pm 0.03$ & $14.78\pm 0.08$ & $13.65\pm 0.17$ & $13.70\pm 0.05$ & $18.48$ & $17.74$ & $17.43$ & $0.216$ \\
$ZOA085829.346-574338.82$ & $275.661$ & $-7.781$ & $0.04$ & $-46.23$ & $2.15$ & $16.26\pm 0.04$ & $15.61\pm 0.03$ & $15.42\pm 0.05$ & $-$ & $-$ & $15.38\pm 0.05$ & $19.50$ & $18.87$ & $18.74$ & $0.216$ \\
$ZOA085851.330-575226.84$ & $275.806$ & $-7.839$ & $0.28$ & $-41.01$ & $7.50$ & $13.90\pm 0.02$ & $13.15\pm 0.02$ & $12.77\pm 0.03$ & $13.31\pm 0.20$ & $12.89\pm 0.07$ & $12.34\pm 0.11$ & $17.50$ & $16.74$ & $16.38$ & $0.235$ \\
$ZOA085826.056-574700.67$ & $275.699$ & $-7.823$ & $0.32$ & $-52.72$ & $2.12$ & $17.80\pm 0.08$ & $16.99\pm 0.06$ & $16.30\pm 0.09$ & $-$ & $-$ & $16.14\pm 0.10$ & $20.74$ & $20.07$ & $19.99$ & $0.235$ \\
\bottomrule
\end{tabular}
\label{tab:irsfcat}
\end{sidewaystable*}

\subsection{Catalogue magnitude completeness limit}
\label{sec:completeness}
\begin{figure}
\begin{center}

    \includegraphics[width=0.33\textwidth]{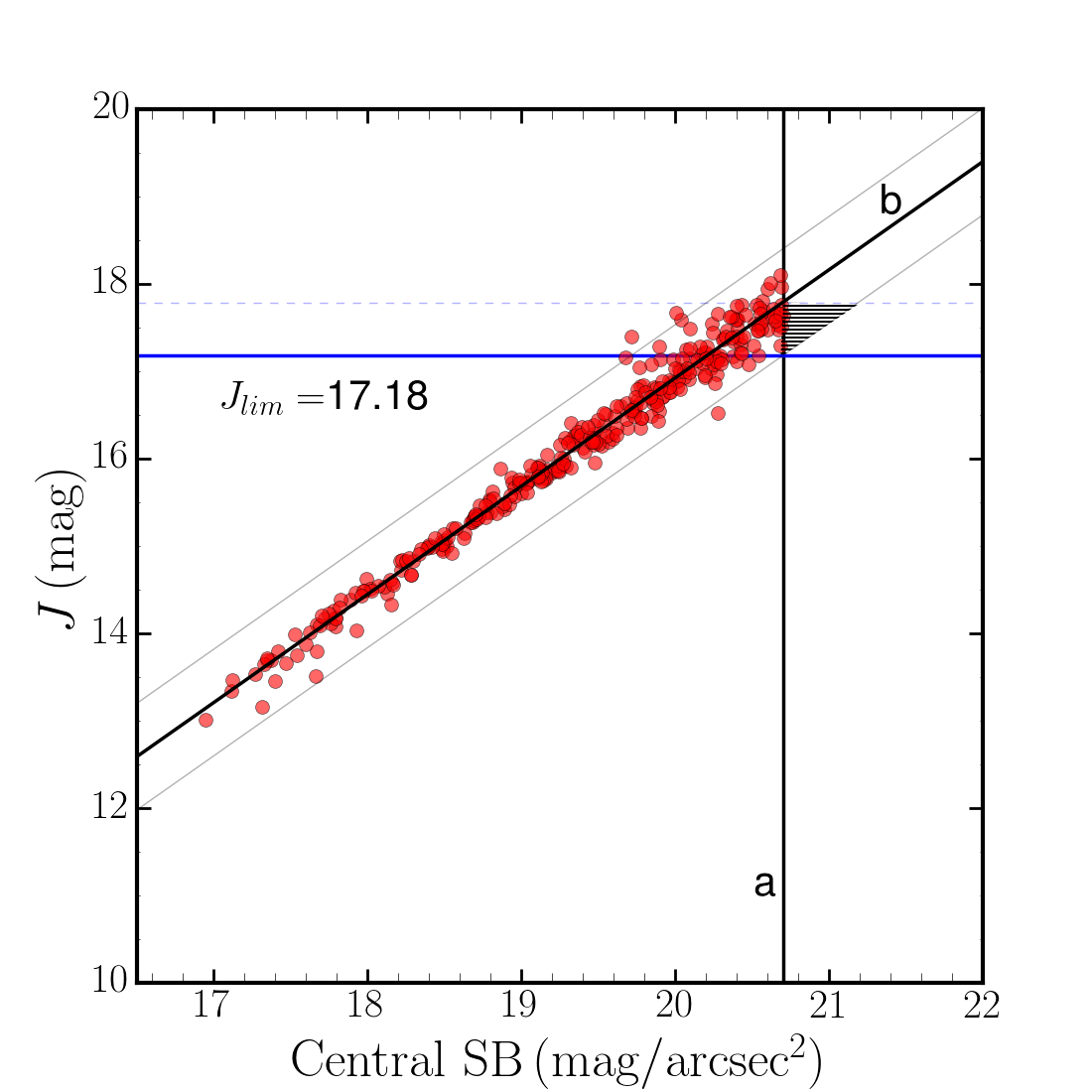}

    \includegraphics[width=0.33\textwidth]{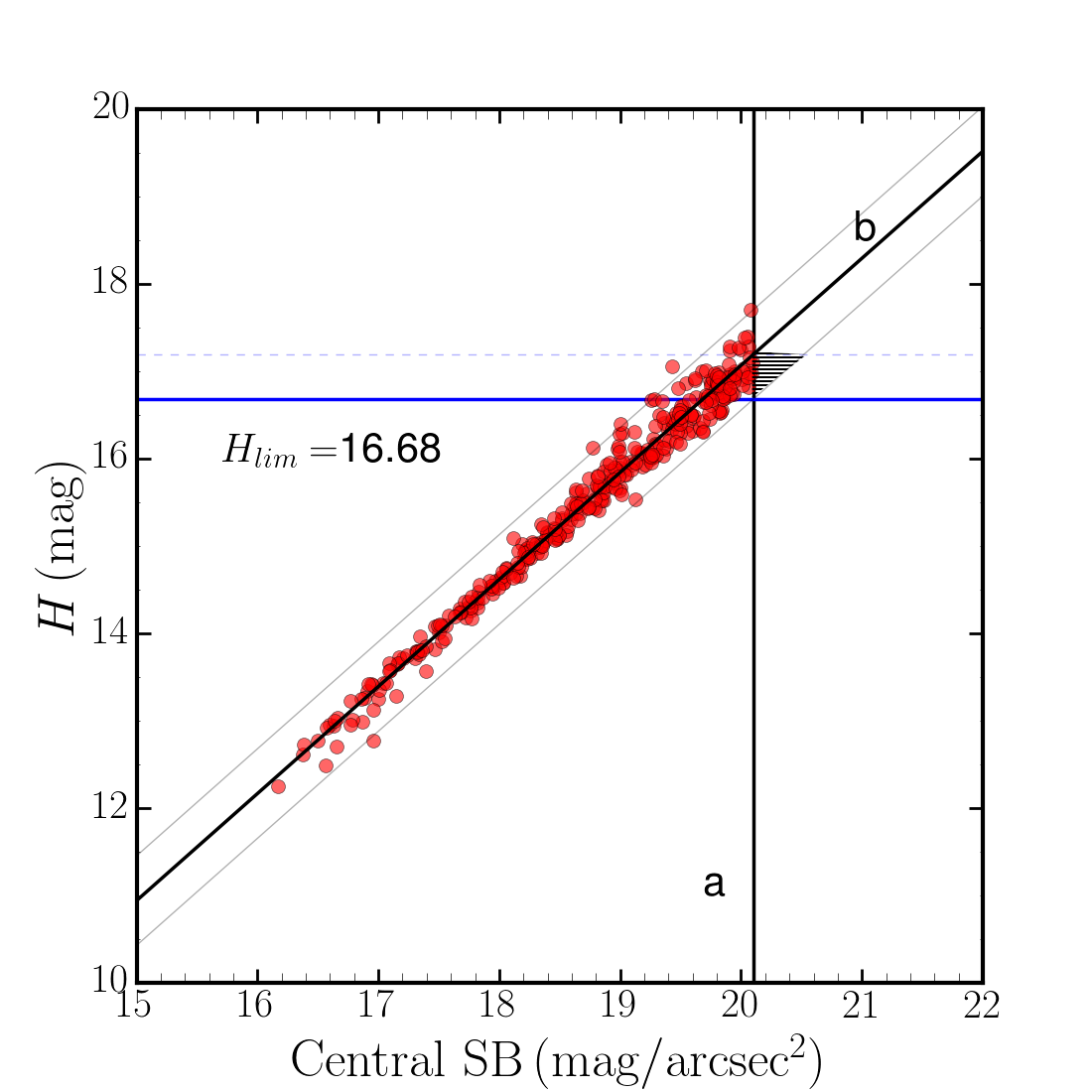}
    
    \includegraphics[width=0.33\textwidth]{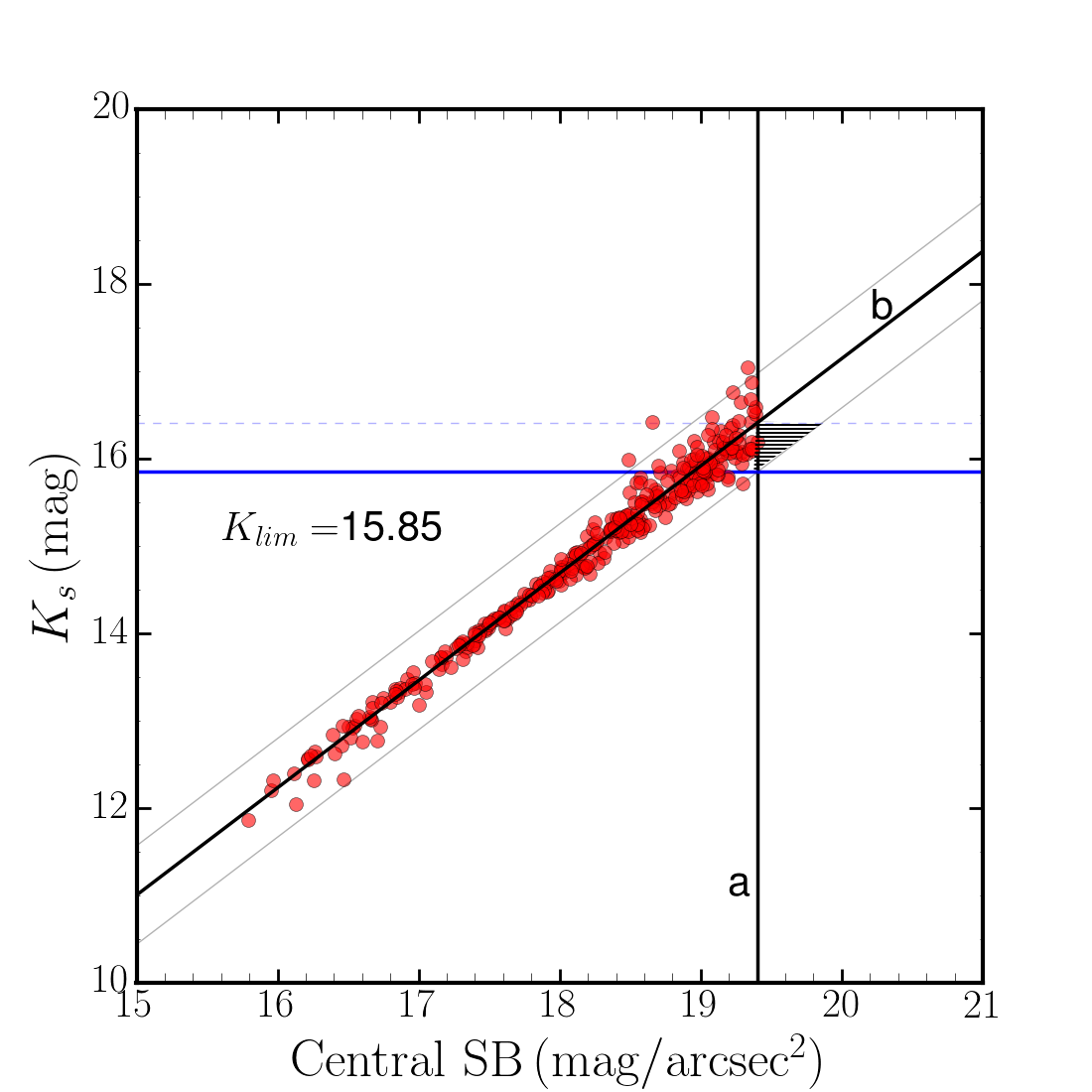}

\caption{Isophotal magnitude vs central surface brightness of candidate galaxies in VC04 in three $J$, $H$ and $K_{s}$-bands. The fitted relation is shown by the solid black line, with the thin gray lines representing a deviation of $3\sigma$ on either side. The completeness magnitude is indicated by the solid thick blue line.} 
\label{fig:magcorr-limit} 
\end{center}
\end{figure}

To perform scientific analysis on the cluster properties a good assessment of the magnitude completeness limits is required. To determine this we applied the procedure described in  \citet{garilli1999} and \citet{andreon2000} which was also followed by \citet{skelton2009} in the analysis of the Norma cluster. This method is based on the assumption that galaxies will not be detected once their surface brightness falls below a certain brightness threshold. In Fig.~\ref{fig:magcorr-limit} the relation between the $K_{s20}$ fiducial isophotal magnitude and the central surface brightnesses of galaxies  in VC04 is shown for the three $J$, $H$ and $K_{s}$-bands.  
Recall that all the galaxies were visually inspected to minimise the number of spurious detections  and the criterion $r>2''$ does not remove real galaxies from the sample (see Section~\ref{sec:sampleanalysis} and Fig.~\ref{fig:colour-radius}). Also note that we measured the mean central surface brightness values within the radius of $3''$, which is larger than the maximum values of the seeing in the images in $J$, $H$ and $K_s$-bands (FWHM$=1\farcs87, 1\farcs76$ and $1\farcs63$ respectively, see Table~\ref{tab:imagestat}), to minimise the effect of PSF-smearing.
An estimate of the difference between $3''$ and $5''$ central surface brightness values is found to be less than $0\fm2$, which is an overestimation for the effect of the highest value of seeing on the central surface brightnesses.

\begin{table}
\caption{Comparison of the completeness limits of VSCL clusters to other NIR surveys performed by IRSF/SIRIUS camera.}
\begin{center}
\begin{threeparttable}
\scalebox{0.95}{
\begin{tabular}{ccccccc}
\toprule
\toprule
  & $J$& $J^{0}$ & $H$& $H^{0}$ & $K_{s}$& $K^{0}_{s}$ \\
 & (mag)&(mag)&(mag)&(mag)&(mag)&(mag) \\
 \midrule
 VC02  &17.31& 17.00 &16.57 & 16.37 &15.78 & 15.65 \\
 VC04  &17.18& 16.98 &16.68& 16.55 &15.85& 15.76 \\ 
 VC05 &17.36 & 17.18 & 16.49 & 16.37 & 15.84 & 15.76 \\
 VC08  & 17.25 & 17.03 & 16.32 & 16.17 & 15.43 & 15.34 \\
 VC10  &17.43 & 17.20 & 16.55 & 16.40 & 15.89 & 15.79 \\
 VC11  & 17.18 & 16.88 & 16.42 & 16.24 & 15.83 & 15.70 \\
 NW\tnote{a} & 16.60& 15.60&15.80& 15.30 &15.40& 14.80\\
 HIZOA \tnote{b} & 16.40& $-$& 15.80& $-$ & 15.40& $-$\\
 Norma \tnote{c} & $-$&$-$& $-$&$-$ & 15.74& $-$\\
 
 \bottomrule
\end{tabular}}
\begin{tablenotes}
\item[a] Norma Wall \citep{riad2010}.\\
\item[b] HIZOA J0836-43 \citep{cluver2008}.\\
\item[c] Norma cluster \citep{skelton2009}.
\end{tablenotes}
\end{threeparttable}
\end{center}
\label{tab:mag-limit}
\end{table}

In Fig.~\ref{fig:magcorr-limit} the identified galaxies are displayed as red circles. The thick black line (b) is the linear fit between the magnitudes and surface brightnesses of the galaxies. The $\pm \, 3\sigma$ deviation are shown as gray lines on either side of the linear fit. The vertical thick black line represents the surface brightness detection limit, which is roughly chosen at where the number of galaxies with faint surface brightnesses starts to decrease and where no data points are found above the fitted line. 
The intersection between the detection limit and the linear fit cannot be taken as the completeness limit because galaxies in the hatched area would not be detected even if they are brighter than the completeness limit. Therefore, where the surface brightness limit line (a) intersects with the $-3\sigma$ of the linear fit was selected to be the magnitude completeness limit. The  equations of the fitted lines and completeness limit values for the VC04 cluster in the $J, H$ and $K_{s}$-bands are as follows: \\ 

$y=(1.23 \pm 0.01) x- (7.85\pm 0.20), \hspace{0.5cm} J_{lim}=17\fm18$ 

$y=(1.22 \pm 0.01) x- (7.44\pm 0.17), \hspace{0.5cm} H_{lim}=16\fm68 $

$y=(1.22 \pm 0.01) x- (7.42\pm 0.18), \hspace{0.5cm} K_{s,lim}=15\fm85$ \\

\noindent
The extinction-corrected values for the magnitude completeness limits were determined by subtracting the maximum values of extinction on galaxies within each cluster (see Table~\ref{tab:foregroundext}) from the observed magnitude completeness values. It was done in the three $J$, $H$ and $K_{s}$-bands. The observed and extinction-corrected completeness magnitude limits for the six VSCL clusters are listed in Table~\ref{tab:mag-limit}. This table also contains the magnitude completeness limits of the other NIR surveys performed with the IRSF/SIRIUS camera. The 2MASS survey completeness magnitude at high Galactic latitudes ($|b|>10\deg$), is limited to $K_{s}=13\fm5$ \citep{skrutskie2006} which shows that the VSCL survey is $\sim 2\fm0$ deeper compared to the 2MASS survey. 
Table~\ref{tab:mag-limit} shows that the observed $K_{s}$-band completeness limit in this work is consistent with the other NIR surveys using the IRSF/SIRIUS camera. Note that the survey area in \citet{riad2010} is located closer to the GP, where the extinction is much higher than this work.

The lowest extinction-corrected value of the $K_s$-band magnitude limit belongs to VC08, $K_s^o=15\fm34$, which after applying the extinction and $k$-correction becomes $K_s^{o,k}=15\fm12$. This value is lower than those of the other five clusters, because the images of VC08 are of slightly lower quality and had slightly higher seeing values than the other cluster images. It is located in a region of higher stellar density resulting in higher sky brightness. The lowest extinction-corrected value of the $K_s$-band magnitude limit for the other five clusters is $K_s^o=15\fm65$, or a  $k$-corrected value of $K_s^{o,k}=15\fm53$.  Therefore, we will use the value of $K_s^{o,k}\sim 15\fm5$ as the magnitude completeness limit in common to all 6 cluster candidates, and their catalogues respectively. At the velocity of the VSCL ($cz\sim 18000$~km~s\textsuperscript{-1}) this translates to $M_{Ks}^o=-21\fm5$.

\section{Structures and richnesses of the six cluster candidates}
\label{sec:sixgalcluster}

Iso-density contour maps and radial density profiles are useful tools for examining the structure and richness of clusters, in the absence of high spectroscopic data coverage. We use this method since only 2MASX galaxies ($K^o_s\la 13\fm5$) in the VSCL region have spectroscopic redshifts, hence they are  $\sim 2\fm0$ brighter than the IRSF photometric data.  We first perform the analysis on four well-studied nearby clusters as a comparison sample for assessment of the six VSCL cluster candidates. 
The selected clusters are Coma, Norma, 3C129 and Virgo. They are all quite local ($v\la 7000~$km~s$^{-1}$) and have NIR ($JHK_s$) photometric data to the approximate same magnitude completeness limit of the more distant VSCL clusters ($17000-22000~$km~s$^{-1}$) and vary in richness from a young, irregular cluster such as Virgo to a virialized, very massive cluster such as Coma.

The properties of these clusters are summarized in Table~\ref{tab:clusterprop}.
We used the 2MASX catalogue \citep{jarrett2000A}, to extract the NIR data for Coma, Norma and Virgo. The NIR data of the 3C129 cluster is obtained from a catalogue of galaxies in the $J$ and $K$ bands provided by \citet{ramatsoku2020}, using the UKIRT Infrared Deep Sky Survey \citep[UKIDSS,][]{lawrence2007}. We performed the analysis on the comparison clusters out to the same radius $r_c<1.5~$Mpc, and completeness magnitude limit $M_{Ks}^o<-21\fm5$ as for the VSCL cluster candidates.

\begin{table*}
\caption{Properties of four comparison clusters, Coma, Norma, 3C129 and Virgo. $\langle A_{Ks} \rangle$ indicates the mean foreground extinction in each cluster derived from \citet{schlafly2011}. The distances are taken from the corresponding citations for the velocity dispersions  ($\sigma_{v}$).}
\begin{center}
\scalebox{0.86}{
\begin{tabular}{ccccccccc}
\toprule
\toprule
Cluster & Distance & X-ray flux  & Virial radius & $\sigma_{v}$ & Virial mass & X-ray mass & $\langle A_{Ks} \rangle$ \\
   & $h^{-1}_{70}$\,Mpc & $10^{-10}$ ergs s\textsuperscript{-1} cm\textsuperscript{-2} &$h^{-1}_{70}$\,Mpc &km s\textsuperscript{-1} &$10^{15}h^{-1}_{70}$\,\MSUN &$10^{15}h^{-1}_{70}$\,\MSUN & mag \\ 
 \midrule
 Coma &99.1 &3.20 \citep{ebeling1998} &2.84 \citep{kubo2007} &1034 \citep{wegner1999} & 2.68 \citep{kubo2007} &1.32 \citep{hughes1989}& 0.003 \\
 Norma &69.2 &2.20 \citep{ebeling2002} &2.14 \citep{Boehringer1996}&925 \citep{woudt2008}& 0.85 \citep{Boehringer1996}  &$1.57$ \citep{Boehringer1996}& 0.070\\
 3C129 & 74.0 & 0.91 \citep{ebeling2002}&1.89 \citep{pinzke2011} &765 \citep{leahy2000}&0.78 \citep{pinzke2011} &0.35 \citep{leahy2000} & 0.350 \\
 Virgo &15.0 &8.21 \citep{ebeling1998} &1.55 \citep{mclaughlin1999} &700 \citep{binggeli1993} &0.65 \citep{kashibadze2020} &0.42 \citep{mclaughlin1999} & 0.009\\

 \bottomrule
\end{tabular}}
\end{center}
\label{tab:clusterprop}
\end{table*}

\begin{figure*}
    \centering
    \includegraphics[width=1.1\textwidth]{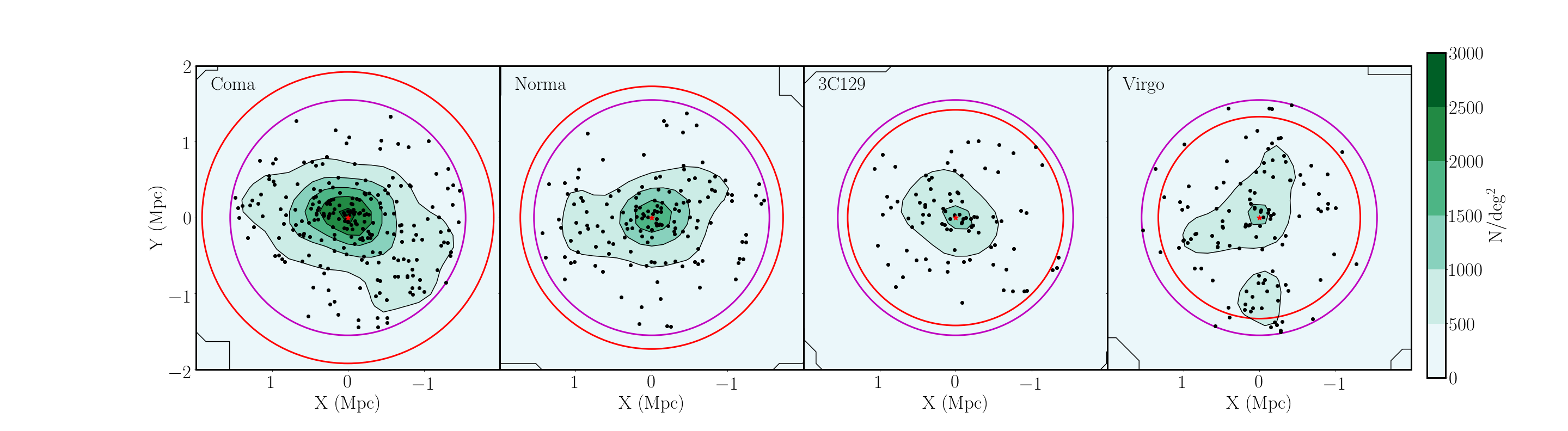}
    \includegraphics[width=1.1\textwidth]{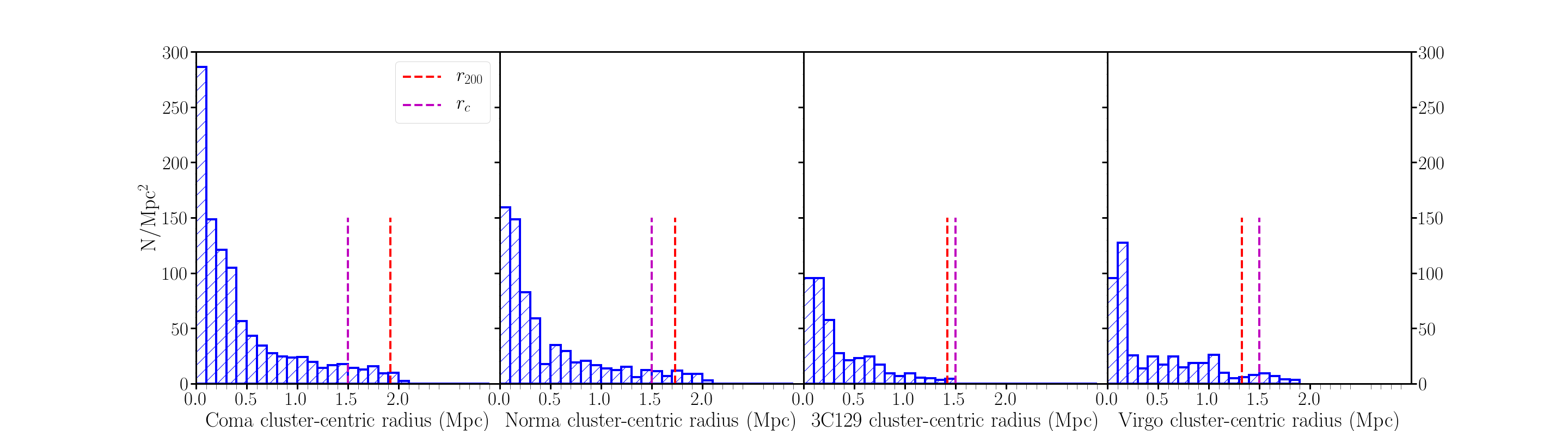}

    \caption{Top panel: the iso-density contour maps of the Coma, Norma, 3C129 and Virgo clusters out to $r_c=1.5\,$Mpc and the extinction-corrected magnitude completeness limit of $M_{Ks}^o<-21\fm5$. The galaxies are shown as black dots and the centres of clusters as red stars. The red and magenta circles represent the $r_{200}$ and $r_c$ respectively. Bottom panel: the radial density profile of the Coma, Norma, 3C129 and Virgo clusters out to $r_c=2\,$Mpc, complete to $M_{Ks}^o<-21\fm5$. The red and magenta lines demonstrate the $r_{200}$ and $r_{c}$ radii respectively. 
    }
    \label{fig:compareclus}
\end{figure*}

To produce the iso-density contour plots, we first determined the number of galaxies per square grid of $\sim 2' \times 2'$. The grids cover the area of each cluster. The number of galaxies in each grid area was converted to a density ($N$/deg$^2$), smoothed by a Gaussian filter to plot the iso-density contour maps. The iso-density contour maps for the four comparison clusters are displayed in the top panel of Fig.~\ref{fig:compareclus}. Galaxies are shown as black dots and the centres of clusters as red stars. The red and magenta circles represent $r_{200}$ and $r_c$ respectively. The $r_{200}$ was derived for each cluster using equation \ref{eq:r200} \citep{carlberg1997}.

\begin{equation}
\begin{aligned}
r_{200}=2.47 \frac{\sigma_{v}}{1000\, \mathrm{km\, s^{-1}}} \frac{1}{\sqrt{\Omega_{\Lambda}+\Omega_{0}(1+z)^3}} h_{70}^{-1}\, \mathrm{Mpc}.
\end{aligned}
\label{eq:r200}
\end{equation}

The number densities of each cluster, shown in the colour bar (see top panel of Fig.~\ref{fig:compareclus}), confirms that the Coma cluster is the richest cluster, followed by Norma, 3C129 and Virgo. This figure also demonstrates an elongated structure in Coma and a clear substructure in the Virgo cluster, while 3C129 shows a symmetric structure in the central region.

\begin{figure*}
\centering
\includegraphics[width=1.19\textwidth]{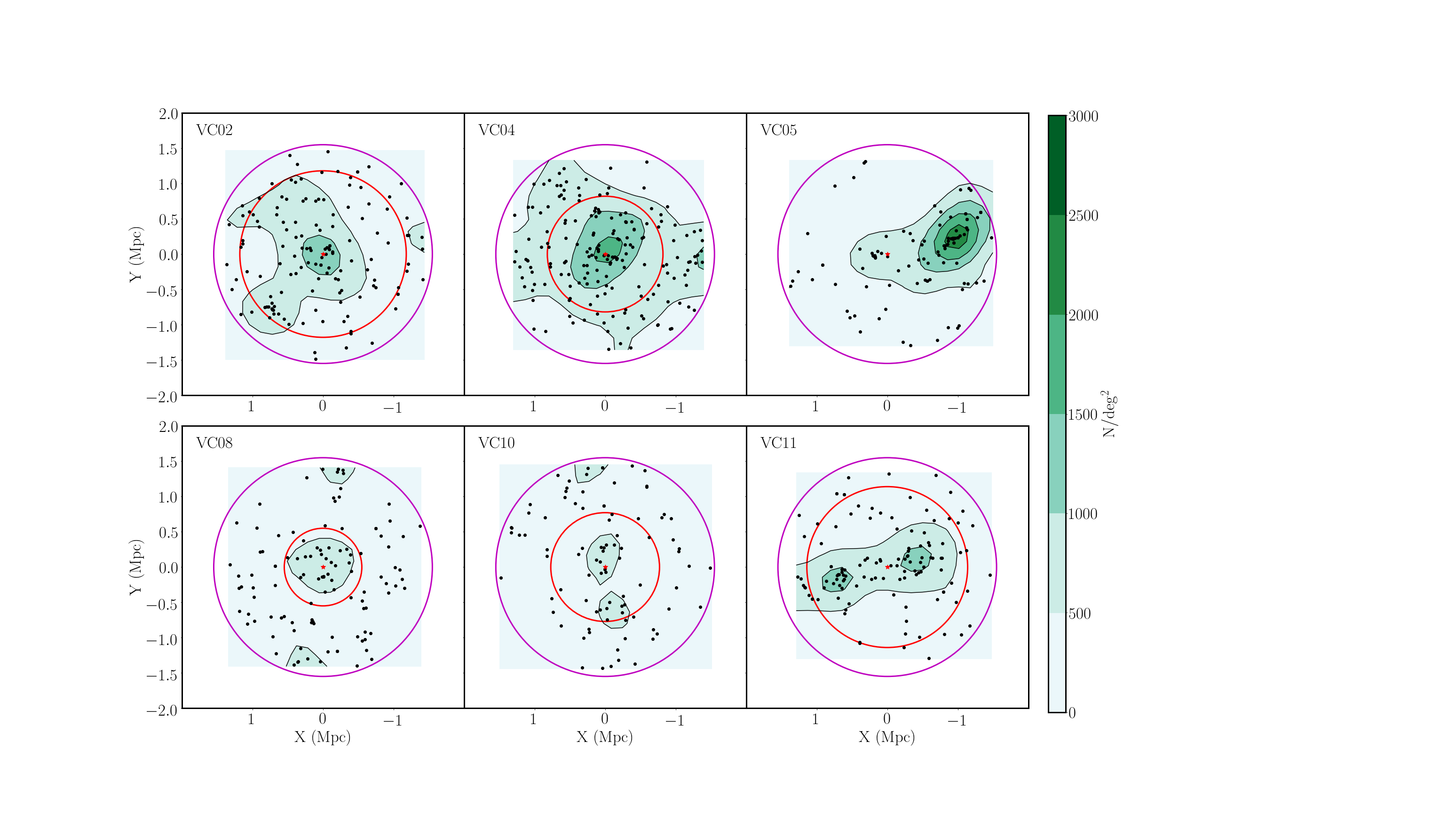}

  \begin{subfigure}[b]{0.29\textwidth}
    \includegraphics[width=\textwidth]{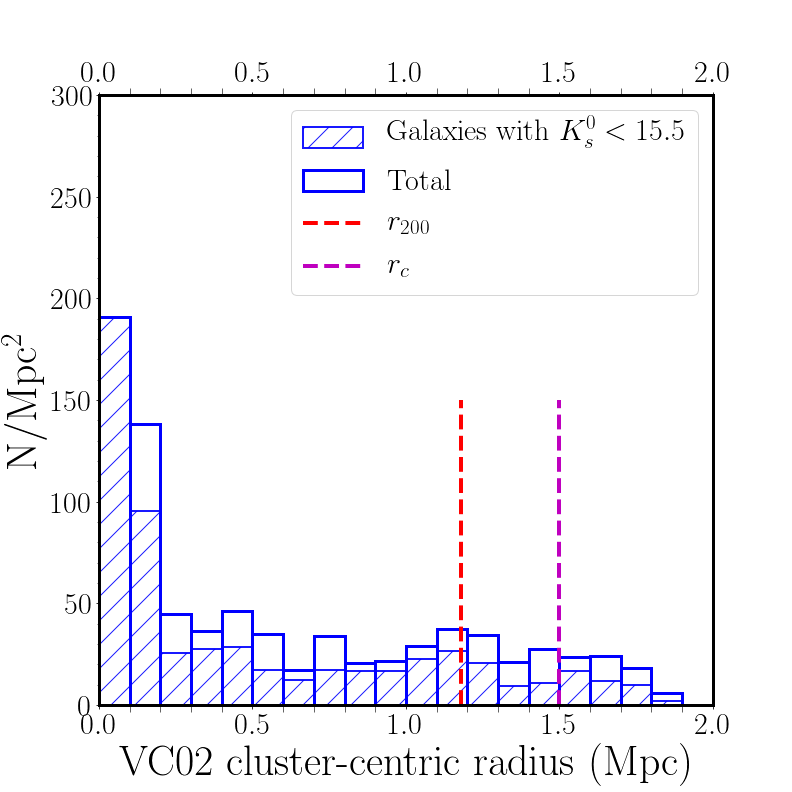}

  \end{subfigure}
  %
  \begin{subfigure}[b]{0.29\textwidth}
    \includegraphics[width=\textwidth]{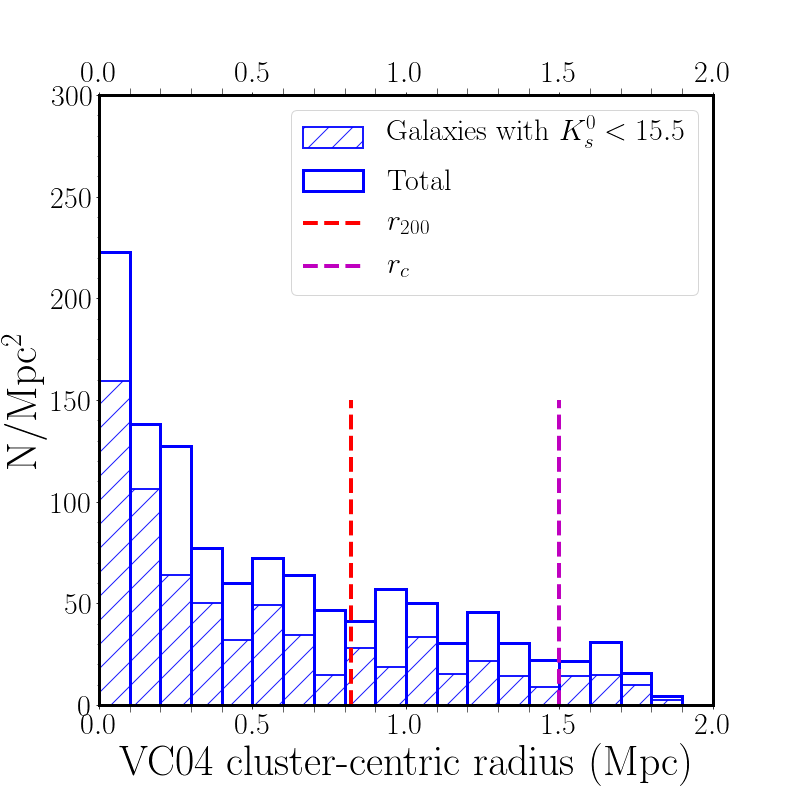}

  \end{subfigure}
  \begin{subfigure}[b]{0.29\textwidth}
    \includegraphics[width=\textwidth]{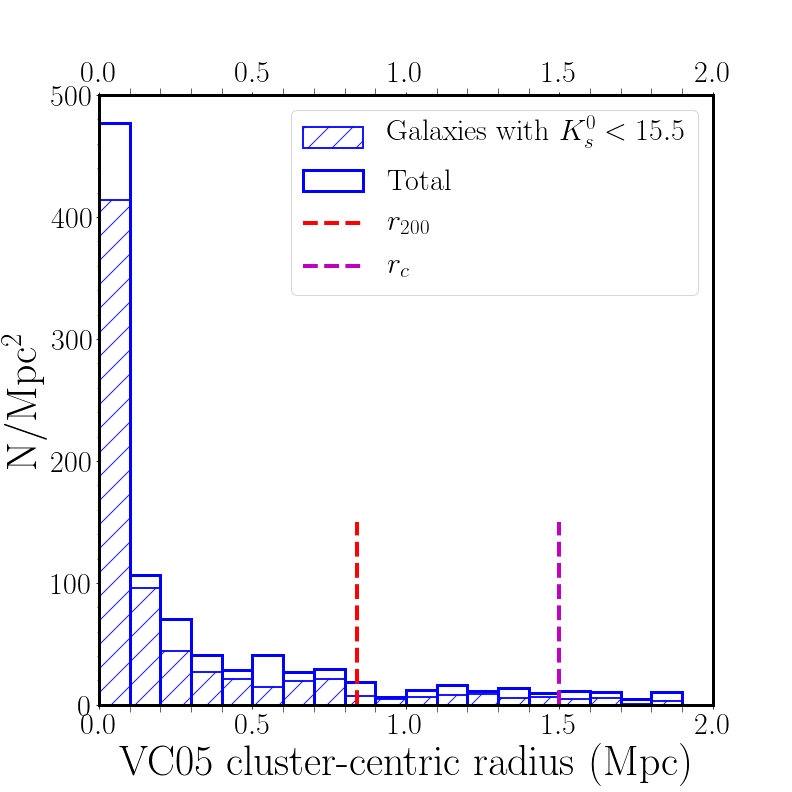}

  \end{subfigure}
  %
  \begin{subfigure}[b]{0.29\textwidth}
    \includegraphics[width=\textwidth]{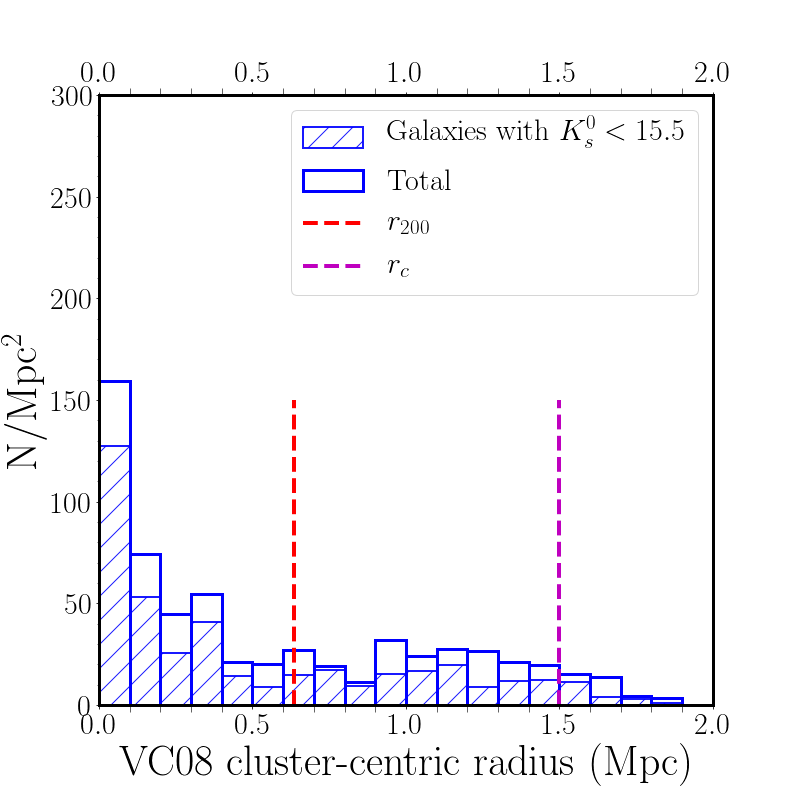}

  \end{subfigure}
  \begin{subfigure}[b]{0.29\textwidth}
    \includegraphics[width=\textwidth]{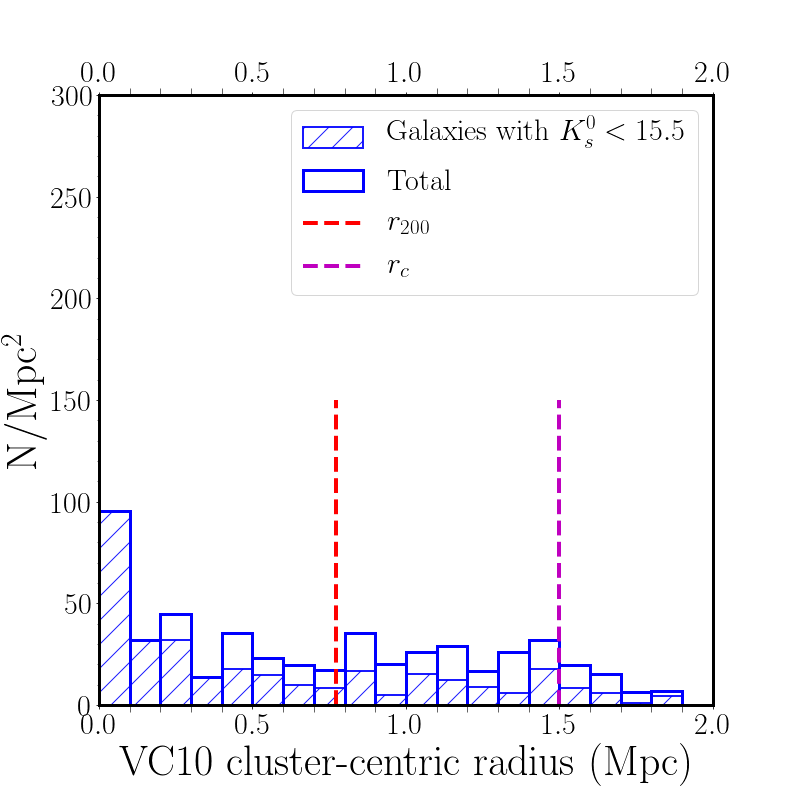}

  \end{subfigure}
  %
  \begin{subfigure}[b]{0.29\textwidth}
    \includegraphics[width=\textwidth]{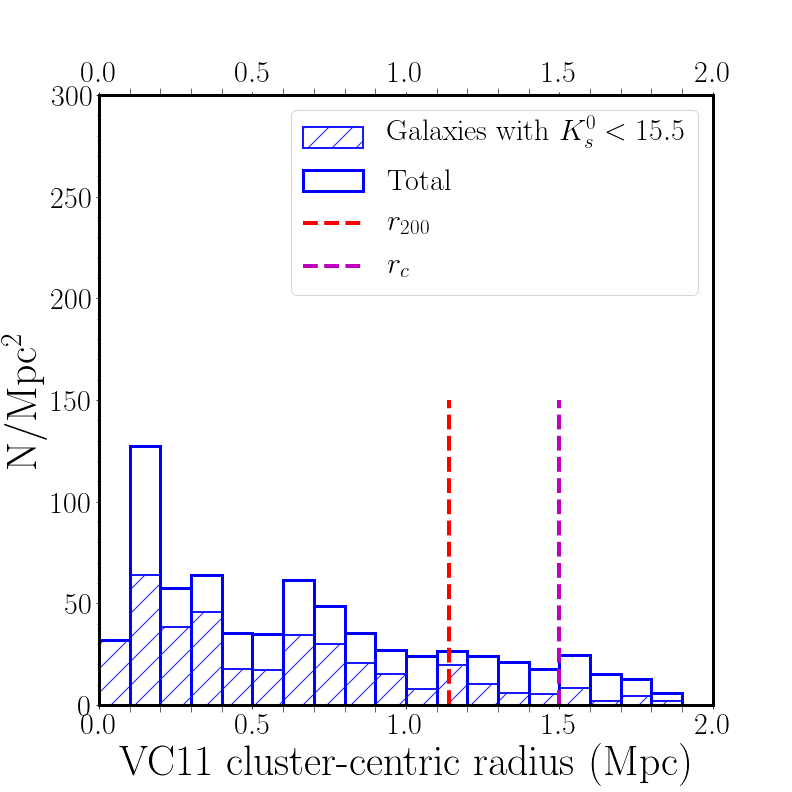}

  \end{subfigure}
 \caption{Top panel: The iso-density contour map of the VSCL clusters to the same density level as the comparison clusters out to $r_c=1.5\,$Mpc, complete to $M_{Ks}^o<-21\fm5$. The galaxies are shown as black dots and the red and magenta circles demonstrate the $r_{200}$ and $r_{c}$ radii respectively. Bottom panel: The radial density profiles of the VSCL clusters out to $r_c=2\,$Mpc. The total number density of galaxies is shown in blue histogram and the completeness-limited sample ($K_{s,lim}^o <15\fm5$) in pink. The red and magenta lines demonstrate the $r_{200}$ and $r_{c}$ radii respectively.} 
 \label{fig:sixvsclradial}
\end{figure*}

To illustrate the degree of concentration of galaxies, we plotted the radial distribution of each cluster, with the number density ($N$/Mpc$^2$) of galaxies in bins of 0.1~Mpc. The radial density profiles of the comparison clusters out to $r_c$ are shown in the bottom panel of Fig.~\ref{fig:compareclus}. The red and magenta dashed lines mark the $r_{200}$ and $r_c$. The radial density profiles of the Coma and Norma clusters show that they are highly concentrated toward their centres and have gradual and smooth distributions which are indicative of rich and massive clusters, while the Virgo cluster shows signs of an irregular cluster \citep{abell1975,bahcall1977}. The 3C129 cluster is an intermediate cluster, not as rich as Coma and Norma, but centrally concentrated.

The top panel of  Fig.~\ref{fig:sixvsclradial} displays the iso-density contour map of the six VSCL cluster candidates, constructed according to the method presented above. To allow a fair comparison of the cluster candidates to the comparison clusters, their iso-density contour maps were determined to the same density levels as those of the Coma, Norma, 3C129 and Virgo clusters. The bottom panel of Fig.~\ref{fig:sixvsclradial} displays the radial density profiles of the VSCL cluster candidates, $N$/Mpc$^2$, out to 2~Mpc.

The analysis shows that VC02 appears not to be as rich as the Coma and Norma clusters but richer than  Virgo or 3C129. However,  foreground extinction is higher for VC02, especially in the right top quadrant, than for any of the other VSCL clusters (see Fig.~\ref{fig:extinction}). Some galaxies might be missing from this analysis due to the heavier obscuration by the dust of the Milky Way. Weak X-ray emission was found for VC02 with offset from the centre towards the north. Furthermore, there is a small clump of galaxies to the south-east of the central region which could be a small group falling into the core of VC02.

VC04 stands out as the richest cluster among the six and is comparable to the Coma and Norma clusters, although its velocity dispersion  seems rather low for a rich cluster ($\sigma_v=455~$km~s$^{-1}$). It is interesting that the observed X-ray emission does  not suggest VC04 to be a rich cluster. The X-ray study of VC04 (B\"ohringer, private communication), using ROSAT \citep{Boehringer1996} in 0.1-2.4~keV, revealed $L_x\sim 5\times 10^{43}$~erg~s$^{-1}$ which would make it more like an irregular Virgo-like cluster. This might well indicate that X-ray emission (in particular in 0.1-2.4~keV) could be underestimated for clusters low in the GP. This could be due to the gas in the Milky Way (hydrogen atoms in the interstellar medium) absorbing the X-ray emission, especially the softest X-ray emission (0.1-2.4~keV).

The centre of VC05 was previously estimated incorrectly based on available redshift data in identifying this cluster at the time \citep{elagali2015}. Using the iso-density contour map, we adjusted the coordinates of the centre of VC05  and list them in Table~\ref{tab:5clustersprop}. Comparison of the iso-density maps indicates that VC05 is much richer than the  3C129 and Virgo clusters and could be comparable to the Norma and Coma clusters. Note that VC05 has the highest central density of all the clusters studied in VSCL. However, due to the previously misidentified centre, the sample of galaxies in VC05 is not complete to $r_c<1.5~$Mpc (and therefore $r_c$ is not indicated in its figure).

Although VC08 is centrally concentrated, it is a relatively poor cluster and not as rich as the 3C129 and Virgo clusters. However, as mentioned earlier, the images of VC08 are of slightly lower quality because of higher seeing values than the other cluster images. VC08 is also located in a region with higher stellar density which results in higher sky brightness.

The iso-density map of VC10 shows that it does not possess a central concentration typical of clusters but rather a similar galaxy density across the observed field indicating that it is more likely some sort of filamentary/wall-like structure rather than  a cluster. This is not surprising in a supercluster, since high density clusters in relatively rich superclusters are connected by lower-density filaments \citep{einasto2007d}.

The iso-density map of VC11 depicts that the previously predicted centre is located in the middle of the two major clumps (or subclusters) in its region. The two clumps are named VC11a and VC11b and their centres are listed in Table~\ref{tab:5clustersprop}. The radial density profiles of each clump show that VC11a (left clump is Fig.~\ref{fig:sixvsclradial}) in more centrally concentrated than VC11b (right clump).  VC11 in general, has the same order of richness as the 3C129 and Virgo clusters. The two subclusters might be merging into a one rich cluster, although more spectroscopic data is needed to confirm this.

\section{Summary and future work}
\label{sec:summary}
To learn more about the VSCL, a series of deep NIR observations of its constituent galaxy clusters were conducted using the IRSF telescope during three observing runs (2015-2019). Each cluster was observed in a mosaic of 32 fields centred on prospective VSCL clusters. Six of the observed clusters have a full set of science quality images with  mean seeing values of FWHM=1\farcs55, 1\farcs46 and 1\farcs37 in the $J$, $H$ and $K_s$-bands respectively. Investigating the effect of foreground extinction in the region of the observed clusters shows that the VC02 cluster has the highest foreground extinction while VC04 and VC05 have the lowest ($\langle A_{Ks} \rangle =0\fm10, 0\fm07$ and $0\fm06$ respectively).

After identifying galaxies and performing photometry on them, we derived the magnitude completeness limit  of the six clusters to be $K_s^{o,k}= 15\fm5$. This value is extinction and $k$-corrected and at the distance of the VSCL translates to $M_{Ks}^o=-21\fm5$, which is $\sim 2\fm0$ deeper than 2MASX. We probed the structures and estimated the richnesses of the six clusters using their iso-density contour maps and radial density profiles and compared them to those of a sample of well-known clusters (Coma, Norma, 3C129 and Virgo). The analysis was performed out to the completeness magnitude and radius of this work, $M_{Ks}^o<-21\fm5$ and $r_c<1.5~$Mpc.
The properties of the six cluster candidates are summarized in Table~\ref{tab:5clustersprop}.

\begin{table*}
\caption{Parameters of the six VSCL clusters.}
\begin{center}
\footnotesize
\begin{threeparttable}
\scalebox{0.95}{
\begin{tabular}{clcccccc}
\toprule
\toprule
  & Cluster & VC02 & VC04 & VC05& VC08 & VC10& VC11\tnote{1} a \& b    \\
 \midrule
 1&RA (J2000) & 08 10 49.20 & 08 36 34.80 &  08 37 46.80 & 08 58 33.60 & 09 49 31.92 &   09 56 52.80, 09 55 29.28 \\
 2&Dec (J2000) & $-49$ 24 36 &  $-55$ 48 54 & $-57$ 25 30 & $-57$ 46 12 & $-43$ 50 42  &   $-43$ 56 24, $-43$ 52 30  \\
 3&$\ell$ (deg) & 264.690 & 272.268 & 273.685 & 275.699 & 271.673&  272.771, 272.532   \\
 4&$b$ (deg) & $-8.608$ & $-8.960$ & $-9.776$ & $-7.801$ &  7.713 &  8.474, 8.369 \\
 5&$N$ &290 & 416 & 279 & 217 & 232 & 281 \\
 6&$N_c$ & 175& 225 & 143 & 127 & 112  & 128 \\
 7&$N_{\mathrm{2MASX}}$ & 43& 82 & 35 & 31 &  29&  48\\ 
 8&$N_r$ & 70 & 85 & 46 & 34 & 26 & 55  \\
 9&Redshift (km~s$^{-1}$)&18955 & 18196 & 21275 & 17298 & 18173 & 17889 \\ 
 10&$\sigma_r$ (km~s$^{-1}$) & 670& 455 & 469 & 355 & 428 & 636 \\ 
 11&$m$ (mag) & 11.35 & 11.85 & 11.93 & 11.60 & 12.21 & 11.76 \\
 12&$A_{Ks}$ (mag) & 0.10 & 0.07 & 0.06 & 0.08 & 0.08 & 0.08 \\
 \midrule
 13&$f$ (\%) & 42 & 38 & 35 & 27 & 25 & 44  \\
 14&$r_{200}$ (Mpc) & 1.18& 0.82 & 0.84 & 0.63 & 0.77 & 1.14 \\
 15&Density (N~Mpc$^{-2}$) & 190& 160 & 413 & 95 & 95 & 254, 95 \\
 16&Richness &intermediate\tnote{2} & very rich\tnote{3} & rich  & poor & poor & intermediate \\
 17&Structure & regular & regular & regular & regular & filament & irregular (two clumps) \\
 18&X-ray & Yes\tnote{4} & Yes\tnote{5} & No & No & No & No \\

 \bottomrule
\end{tabular}}
\begin{tablenotes}
\item[1] VC11 consists of two main clumps (subclusters), VC11a and VC11b.\\
\item[2] 3C129- or Virgo-like richness.\\
\item[3] Coma- or Norma-like richness.\\
\item[4] $L_x\sim 2.6\times 10^{43}$~erg~s$^{-1}$, $\sim 19\farcm50$ offset from the centre.\\
\item[5] $L_x\sim 5\times 10^{43}$~erg~s$^{-1}$ $\sim 30\farcs34$ offset from the centre.
\end{tablenotes}
\end{threeparttable}
\end{center}
\label{tab:5clustersprop}
\end{table*}

The parameters of Table~\ref{tab:5clustersprop} are defined as follows:\\
\noindent {\sl Row }$1$: Equatorial  longitude; (RA) [hhmmss.ss].\\
\noindent {\sl Row }$2$: Equatorial  latitude; (Dec) [ddmmss].\\
\noindent {\sl Row }$3$: Galactic  longitude; ($\ell$) [deg].\\
\noindent {\sl Row }$4$: Galactic  latitude; ($b$) [deg].\\
\noindent {\sl Row }$5$: Total number of identified galaxies; ($N$).\\
\noindent {\sl Row }$6$: Number of identified galaxies brighter than the completeness magnitude limit ($K^o_{s,lim}<15\fm5$); ($N_c$).\\
\noindent {\sl Row }$7$: Number of 2MASX galaxies within the cluster region; ($N_{2MASX}$).\\
\noindent {\sl Row }$8$: Number of galaxies with redshift, brighter than $K^o_{s,lim}<15\fm5$; ($N_r$).\\
\noindent {\sl Row }$9$: Mean $v_{hel}$; (Redshift) [\kms].\\
\noindent {\sl Row }$10$: Velocity dispersion; ($\sigma_{r}$) [\kms].\\
\noindent {\sl Row }$11$: Magnitude of the brightest cluster galaxy; ($m$) [mag].\\
\noindent {\sl Row }$12$: Mean value of the foreground extinction in the region of clusters; ($A_Ks$) [mag].\\
\noindent {\sl Row }$13$: Fraction of galaxies with redshift up to $K^o_{s,lim}<15\fm5$ and within $r_c<1.5~$Mpc; ($f$).\\
\noindent {\sl Row }$14$: $r_{200}$ [Mpc] (see equation \ref{eq:r200}).\\
\noindent {\sl Row }$15$: Central number density within 0.1~Mpc; (Density) [N~Mpc$^{-2}$].\\
\noindent {\sl Row }$16$: Estimated richness of the cluster based on studying the velocity dispersion, iso-density maps and radial density plots of clusters and comparing them with those of the comparison clusters; (Richness).\\
\noindent {\sl Row }$17$: Structure of the clusters based on the iso-density map and the shape of the radial density profile; (Structure).\\
\noindent {\sl Row }$18$: X-ray emission in 0.1-2.4~keV, using ROSAT (B\"ohringer, private communication); (X-ray).
   
Our analysis shows that VC04 is the richest of the six. It is a regular and massive cluster comparable to Coma and Norma, although its velocity dispersion seems rather low for a rich cluster. VC02 and VC05 are relatively rich clusters while VC08 was found to be a poor cluster. VC05 has the highest central number density among the six clusters.   
VC10 has a filament-like structure and VC11 is an intermediate cluster consisting of two subclusters.  
It appears that many of the VSCL clusters (VC02, VC04, VC05 and VC11) are not relaxed yet and are still evolving. For two of the six clusters X-ray emission in the ROSAT band (0.1-2.4~keV) has been detected, a Virgo-like emission for the VC04 cluster very close to its estimated centre, and a weak X-ray emission for VC02 with an offset from its centre. The X-ray emissions are not consistent with the richnesses of VC02 and VC04 based on what was found in this analysis and we suspect they might have been underestimated due to obscuration by Galactic foreground gas.

The next paper of the series will be dedicated to the properties of VC04. We will derive and analyse its luminosity function and use SALT spectroscopy (in hand) to revise its velocity dispersion, perform a dynamical analysis and estimate its mass. 
We also plan to investigate the \HI~deficiency of spiral galaxies in VC04 using the MeerKAT-16 \citep{jonas2016} observations.  
These observations will help to further characterise VC04 as a rich cluster.

\section*{Acknowledgements}

We would like to acknowledge the support from the Science Faculty at University of Cape Town (UCT), the South African Research Chairs Initiative of the Department of Science and Technology and the South African National Research Foundation. This paper uses observations made at the South African Astronomical Observatory (SAAO). This publication makes use of data products from the Two Micron All Sky Survey, which is a joint project of the University of Massachusetts and the Infrared Processing and Analysis Center/California Institute of Technology, funded by the National Aeronautics and Space Administration and the National Science Foundation. We would like to thank T. Jarrett for the fruitful discussions and the guidance. We thank the IRSF team and Japanese collaborators, T. Nagayama, K. Morihana, Y. Nakajima, N. Tetsuya for assisting with the IRSF observations and software. K. Said is supported by the Australian Government through the Australian Research Council’s Laureate Fellowship funding scheme (project FL180100168).

\section*{Data Availability}

The data underlying this article are available in its online supplementary material.



\bibliographystyle{mnras}
\bibliography{references}



\appendix

\section{Source Extractor configuration file}

\begin{table*}
\caption{Source Extractor parameters}
\centering
\begin{tabular}{lcl}
\toprule
\toprule
 Parameter & Value& Description\\
 \midrule
 DETECT\_MINAREA & 10&  minimum number of pixels above threshold  \\
 DETECT\_THRESH & 1.5&  <sigmas> or <threshold>, <ZP> in mag arcsec\textsuperscript{-2}\\
 ANALYSIS\_THRESH & 1.5&  <sigmas> or <threshold>, <ZP> in mag arcsec\textsuperscript{-2}\\
 FILTER & Y&  apply filter for detection (Y or N)?\\
 FILTER\_NAME & default.conv&  name of the file containing the filter \\
 DEBLEND\_NTHRESH & 32&  Number of deblending sub-thresholds \\
 DEBLEND\_MINCONT & 0.001 &  Minimum contrast parameter for deblending\\
 CLEAN & Y&  Clean spurious detections? (Y or N)?\\
 CLEAN\_PARAM & 1.0&  Cleaning efficiency\\
 MASK\_TYPE & CORRECT &  Type of detection MASKing: can be NONE, BLANK or CORRECT\\
 PHOT\_APERTURES & 5 &  MAG\_APER aperture diameter(s) in pixels \\
 PHOT\_AUTOPARAMS & 2.5, 3.5 &  MAG\_AUTO parameters: <Kron\_fact>, <min\_radius> \\
 PHOT\_PETROPARAMS & 2.0, 3.5&  MAG\_PETRO parameters: <Petrosian\_fact>, <min\_radius>\\
 SATUR\_LEVEL & 50000.0&  level (in ADUs) at which arises saturation\\
 MAG\_ZEROPOINT & zero-point&  magnitude zero-point \\
 GAIN & 91.66&  detector gain in e-/ADU \\
 GAIN\_KEY & GAIN&  keyword for detector gain in e-/ADU\\
 PIXEL\_SCALE & 0.45&  size of pixel in arcsec (0=use FITS WCS info) \\
 SEEING\_FWHM & seeing&  stellar FWHM in arcsec \\
 STARNNW\_NAME & default.nnw&  Neural-Network\_Weight table filename \\
 BACK\_SIZE & 64&  Background mesh: <size> or <width>, <height>\\
 BACK\_FILTERSIZE & 3&  Background filter: <size> or <width>, <height> \\
 BACKPHOTO\_TYPE & GLOBAL &  can be GLOBAL or LOCAL \\
 \bottomrule
\end{tabular}
\label{tab:Source Extractor}
\end{table*}


\bsp	
\label{lastpage}
\end{document}

%% file: abb.tex
\def\la{\mathrel{\hbox{\rlap{\hbox{\lower4pt\hbox{$\sim$}}}\hbox{$<$}}}}
\def\ga{\mathrel{\hbox{\rlap{\hbox{\lower4pt\hbox{$\sim$}}}\hbox{$>$}}}}

\def\deg{{^\circ}}

\def\fm{\hbox{$.\!\!^{\rm m}$}}

\def\fdg{\hbox{$.\!\!^\circ$}}
\def\farcm{\hbox{$.\mkern-4mu^\prime$}}
\def\farcs{\hbox{$.\!\!^{\prime\prime}$}}

\newcommand{\MSUN}{${\rm M}_\odot$}

\newcommand{\kms}{{\,km\,s$^{-1}$}}

\newcommand{\HI}{\mbox{\normalsize H\thinspace\footnotesize I}}

\def\apj    {{ApJ }}
\def\apjl   {{ApJL }}

\def\aj     {{AJ }}
\def\aap    {{A\&A }}

\def\aaps   {{A\&AS }}

\def\araa   {{ARA\&A }}
\def\aarv   {{A\&ARv }}

\def\mnras  {{MNRAS }}

\def\pasa   {{PASA }}
\def\pasp   {{PASP }}

\def\prd    {{PhRvD }}

\def\procspie {{Proc. SPIE }}